\documentclass[showpacs,onecolumn,superscriptaddress]{revtex4}
\usepackage{amsmath,graphicx}
\usepackage[next]{inputenc}
\usepackage[dvips]{epsfig}
\usepackage{hyperref}
\def\be{\begin{equation}}
\def\ee{\end{equation}}

\def\bi{\begin{itemize}}
\def\ei{\end{itemize}}
\def\bn{\begin{enumerate}}
\def\en{\end{enumerate}}
\def\bea{\begin{eqnarray}}
\def\eea{\end{eqnarray}}
\def\no{\nonumber}
\def\ba{\begin{array}}
\def\ea{\end{array}}
\def\bd{\begin{displaymath}}
\def\ed{\end{displaymath}}

\begin{document}

\title{Quench dynamics and ground state fidelity of the one-dimensional extended
quantum compass model in a transverse field}
\author{R. Jafari}
\affiliation{Asia Pacific Center for Theoretical Physics (APCTP), Pohang, Gyeongbuk, 790-784, Korea}
\email[]{jafari@apctp.org, rohollah.jafari@iasbs.ac.ir}
\affiliation{Department of Physics, Institute for Advanced
Studies in Basic Sciences (IASBS), Zanjan 45137-66731, Iran}

\begin{abstract}
We study the ground state fidelity, fidelity susceptibility and quench
dynamics of the extended quantum compass model in a transverse field.
This model reveals a rich phase diagram which includes several critical surfaces depending
on exchange couplings. We present a characterization of quantum phase transitions in terms
of the ground state fidelity between two ground states obtained for two different values of
external parameters. However, we derive scaling relations describing the singular behavior
of fidelity susceptibility in the quantum critical surfaces. Moreover, we study the time
evolution of the system after a critical quantum quench using the Loschmidt. We find that the revival
times of Loschmidt echo are given by $T_{rev}=N/2v_{max}$, where $N$ is the size of
the system and $v_{max}$ is the maximum of lower bound group velocity of quasi-particles.
Although the fidelity susceptibility shows the same exponent in all critical surfaces,
the structure of the revivals after critical quantum quenches displays two different regimes
reflecting different equilibration dynamics.
\end{abstract}
\date{\today}


\maketitle

\section{Introduction \label{introduction}}

In the last few years a big effort has been assigned to the analysis of
quantum phase transitions (QPTs) from the perspective of quantum information \cite{Osterloh,Vidal,Osborne,kargarian1,kargarian2,Jafari1,kargarian3,Jafari2,Langari,Jafari,Mehran,Jafari5}.
Entanglement and fidelity have been accepted as new notions to
characterize quantum phase transitions. Entanglement, referring to
quantum correlations between subsystems is a good indicator of quantum
phase transitions, because the correlation length diverges at the quantum
critical points \cite{Osterloh,Vidal,Osborne}.
The fidelity which is a measure of distance between quantum states,
could be also a nice tool to study the drastic change in the ground states
in quantum phase transitions \cite{Zanardi}.

On the other hand, recent advances in the studies of ultra-cold atoms trapped
in optical lattices introduced a new tool to simulate the
dynamics of interacting quantum many-body systems
in non-equilibrium strongly correlated quantum systems
\cite{Greiner,Jaksch}. These new opportunities are concentrated by the pioneering
experiments on tunable Mott insulator to superfluid quantum phase transitions,
observed by utilization of the optical lattice potential in three-dimensional
3D \cite{Greiner} and 1D \cite{Stoferle} systems.

Recently the investigation of non-equilibrium properties of closed quantum systems have
been getting a lot of attention for several reasons. Specifically it has been applied to quantum
information in which decoherence and entanglement dynamics play an essential
role. From the theoretical point of view, it is very important to understand
the notion of universality for a system away
from equilibrium, where the traditional concepts of phase,
fixed point and renormalization group fail.
Driving a system out of equilibrium is done
in many ways. Most of the attention has been focused on quantum
quenches \cite{Chandra}, namely, sudden changes of
the external parameters of the Hamiltonian controlling the
unitary evolution of the closed system. One of the conventional methods of
understanding the dynamics of a system after a quench is the
Loschmidt echo (LE), which is a benchmark of the partial or full reappearance
of the original state as a function of time \cite{Gorin,Jacquod}.
The LE is defined as follows: if a quantum state $|\psi\rangle$ evolves with two
Hamiltonians $H$ and $H'$, respectively, the LE is the measure of
the overlap given by $LE=|\langle\psi|\exp(\imath H't)\exp(-\imath Ht)|\psi\rangle|^{2}$.
If the system admits the ground state, LE is a dynamical version of the ground state fidelity.
Recently, the time behavior of the LE has been studied
in well-known models, in particular the $XY$ spin chain \cite{Venuti2,Quan} and
cluster XY chain \cite{Montes}. High values of the LE mean that the system is
approaching the initial state. Typically, the LE will decay
exponentially at first and then start oscillating around an
average value \cite{Venuti2}. If the system is finite,
the time evolution is quasi-periodic, forcing the system
arbitrarily close to the initial state for long enough times. The
system will show revivals, i.e., times when the value
of the LE is greater than the average value. The structure of these revivals may be greatly influenced
by criticality \cite{Venuti2}.
In this work, the phase diagram and universality of the one-dimensional extended quantum
compass model (EQCM) \cite{Eriksson,Mahdavifar,Motamedifar} in a transverse filed \cite{Jafari3,Jafari4}
will be studied by means of the fidelity of the ground state and fidelity susceptibility.
This inhomogeneous model covers a group of well-known spin models as its special cases and shows a
rich phase diagram. It should be mentioned that the phase diagram \cite{Motamedifar},
quantum correlation \cite{You}, bipartite entanglement \cite{Liu} and
fidelity \cite{Motamedifar2}, of this model have been studied numerically using the exact
diagonalization method and infinite time-evolving block decimation \cite{Vidal2}.
This study is an important addition to the literature as to the best of our knowledge, the
quenches of a Hamiltonian which also undergoes a very unique type of quantum phase transition \cite{Eriksson,Jafari3,Jafari4},
has not been investigated before.
However, it will be instructive to study the quenches near quantum critical points because of
the expected universality of the response of the system, and thus the possibility of using the quench
dynamics as a nonequilibrium probe of phase transitions. Then, we will explore the quenching dynamics of
this model, when the transverse field or the exchange couplings is quenched.

\section{The Hamiltonian and its Exact Solution\label{EQCMTF}}

Consider the Hamiltonian

\bea
\label{eq1}
H=\sum_{n=1}^{N'}[J_{1}\sigma^{x}_{2n-1}\sigma^{x}_{2n}+
J_{2}\sigma^{y}_{2n-1}\sigma^{y}_{2n}+ L_{1}\sigma^{x}_{2n}\sigma^{x}_{2n+1}
+L_{2}\sigma^{y}_{2n}\sigma^{y}_{2n+1}+h(\sigma^{z}_{2n-1}+\sigma^{z}_{2n})].
\eea
where $J_{1}$ and $J_{2}$ are the odd bonds exchange couplings, $L_{1}$ and $L_{2}$
are the even bond exchange couplings and $N=2N'$ is the number of spins.
It should be pointed out that although periodic and antiperiodic boundary
conditions differ in \textit{O(1/N)} terms, this difference usually does not
affect the phase diagram or other quantities in the thermodynamic limit.
However, although it can be important in the LE that
is typically exponentially small in $N$ and the boundary conditions could have a dramatic
effect in the case of the critical quench \cite{Happola}, but for simplicity we assume periodic
boundary conditions.
This model embraces a group of the other familiar spin models as its special
cases, such as the quantum Ising model in a transverse field for $J_{2}=L_{2}=0$,
the transverse field XY model for $J_{1}=L_{1}$ and $J_{2}=L_{2}$, and the transverse
field XX model for $J_{1}=J_{2}=L_{1}=L_{2}$.
The above Hamiltonian [Eq. (\ref{eq1})] can be exactly diagonalized by
standard Jordan-Wigner transformation \cite{Jordan,Jafari3,Jafari4}
as defined below,

\bea
\no
\sigma^{x}_{j}=b^{+}_{j}+b^{-}_{j},~~
\sigma^{y}_{j}=b^{+}_{j}-b^{-}_{j},~~
\sigma^{z}_{j}=2b^{+}_{j}b^{-}_{j}-1,~~
b^{+}_{j}=c^{\dag}_{j}~e^{i\pi\Sigma_{m=1}^{j-1}c^{\dag}_{m}c_{m}},~~
b^{-}_{j}=e^{-i\pi\Sigma_{m=1}^{j-1}c^{\dag}_{m}c_{m}}~c_{j}
\eea

which transforms spins into fermion operators $c_{j}$.

The crucial step is to define independent Majorana fermions
\cite{Perk,Sengupta} at site $n$, $c_{n}^{q}\equiv c_{2n-1}$
and $c_{n}^{p}\equiv c_{2n}$. This can be regarded as
quasiparticles' spin or as splitting the chain into bi-atomic
elementary cells \cite{Perk}.

Substituting for $\sigma^{x}_{j}$, $\sigma^{y}_{j}$ and
$\sigma^{z}_{j}$ ($j=2n, 2n-1$) in terms of Majorana
fermions followed by a Fourier transformation,
Hamiltonian Eq.(\ref{eq1}) (apart from an additive constant), can be written as

\bea
\no
H^{+}=\sum_{k}\Big[Jc_{k}^{q\dag}c_{-k}^{p\dag}+Lc_{k}^{q\dag}c_{k}^{p}+
2h(c_{k}^{q\dag}c_{k}^{q}+c_{k}^{p\dag}c_{k}^{p})+h.c.\Big],
\eea
where $J=(J_{1}-J_{2})-(L_{1}-L_{2})e^{ik}$, $L=(J_{1}+J_{2})+(L_{1}+L_{2})e^{ik}$
and $k=\pm\frac{j\pi}{N'},~(j=1,3,\cdots,N'-1)$.\\

By grouping together terms with $k$ and $-k$, the Hamiltonian is transformed into a sum
of independent terms acting in the 4-dimensional Hilbert spaces generated by $k$ and
$-k$ ($H^{+}=\bigoplus_{k>0}H_{k}^{+})$, in other word $[H_{k},H_{k'}]=0$ in which
\bea
\label{eq2}
H_{k}&=&(Jc_{k}^{q\dag}c_{-k}^{p\dag}+Lc_{k}^{q\dag}c_{k}^{p}
-J^{\ast}c_{k}^{q}c_{-k}^{p}-L^{\ast}c_{k}^{q}c_{k}^{p\dag}
+Jc_{-k}^{q\dag}c_{k}^{p\dag}+Lc_{-k}^{q\dag}c_{-k}^{p}
-J^{\ast}c_{-k}^{q}c_{k}^{p}-L^{\ast}c_{-k}^{q}c_{-k}^{p\dag})\\
\no
&+&2h(c_{k}^{q\dag}c_{k}^{q}+c_{k}^{p\dag}c_{k}^{p}+c_{-k}^{q\dag}c_{-k}^{q}
+c_{-k}^{p\dag}c_{-k}^{p}).
\eea

Hamiltonian Eq. (\ref{eq2}) can be written in the diagonal block form

\bea
\label{eq3}
H=\sum_{k}\Gamma^{\dag}_{k}.A(k).\Gamma_{k}
\eea

where $\Gamma^{\dag}_{k}=(c_{k}^{q},c_{-k}^{q\dag},c_{k}^{p},c_{-k}^{p\dag})$ and
\bea
\no
A(k)=\left(
       \begin{array}{cccc}
         2h & 0 & L & J \\
         0 & -2h & -J & -L \\
         L^{\ast} & -J^{\ast} & 2h & 0 \\
         J^{\ast} & -L^{\ast} & 0 & -2h \\
       \end{array}
     \right).
\eea

By using the element of the new vector $\Gamma'^{\dag}_{k}=(\gamma_{k}^{q},\gamma_{-k}^{q\dag},
\gamma_{k}^{p},\gamma_{-k}^{p\dag})$
which could be described by unitary transformation $\Gamma'_{k}=U_{k}\Gamma_{k}$ (see Appendix
A), the matrix $A(k)$ can be diagonalized easily and we find the Hamiltonian
Eq. (\ref{eq2}) in a diagonal form.

\bea
\label{eq4}
H=\sum_{k}\Big[E^{q}_{k}(\gamma_{k}^{q\dag}\gamma_{k}^{q}-\frac{1}{2})+
E^{p}_{k}(\gamma_{k}^{p\dag}\gamma_{k}^{p}-\frac{1}{2})\Big],
\eea
where $E^{q}_{k}=\sqrt{a+\sqrt b}$ and $E^{p}_{k}=\sqrt{a-\sqrt b}$, in which

\bea
\no
a=4h^{2}+|J|^{2}+|L|^{2},~~
b=(16h^{2}+2|J|^{2})|L|^{2}+J^{2}{L^\ast}^{2}+{J^\ast}^{2}L^{2}
\eea

The ground state ($E_{0}$) and the first excited state ($E_{1}$) energies are obtained from Eq.(\ref{eq3}),

\bea
\no
E_{0}=-\frac{1}{2}\sum_{k}(E^{q}_{k}+E^{p}_{k}),~~E_{1}=-\frac{1}{2}\sum_{k}(E^{q}_{k}-E^{p}_{k}).
\eea

It is straightforward to show that the energy gap vanishes at $h_{0}=\sqrt{(J_{1}+L_{2})(J_{2}+L_{1})}$
and $h_{\pi}=\sqrt{(J_{1}+L_{2})(J_{2}-L_{1})}$ in the thermodynamic limit.

So, the quantum phase transition which could be driven by the transverse-field
depending on exchange couplings, occurs at $h_{0}$ and $h_{\pi}$.

By a rather lengthy calculation on the unitary transformation we can obtain the whole spectrum
and the eigenstate of the Hamiltonian which have been written in the vacuum $k$th
mode of $c_{k}^{q}$ and $c_{k}^{p}$,

\bea
\no
|\psi_{m}\rangle&&=\prod_{k}[v_{1}^{m}|0\rangle+v_{2}^{m}~c_{k}^{{q\dag}}c_{-k}^{{q\dag}}|0\rangle+
v_{3}^{m}~c_{k}^{{q\dag}}c_{-k}^{{p\dag}}|0\rangle
+v_{4}^{m}~c_{-k}^{{q\dag}}c_{k}^{{p\dag}}|0\rangle+v_{5}^{m}~c_{k}^{{p\dag}}c_{-k}^{{p\dag}}|0\rangle
+v_{6}^{m}~c_{k}^{{q\dag}}c_{k}^{{p\dag}}|0\rangle\\
\label{eq5}
&&+v_{7}^{m}~c_{-k}^{{q\dag}}c_{-k}^{{p\dag}}|0\rangle
+v_{8}^{m}~c_{k}^{{q\dag}}c_{-k}^{{q\dag}}c_{k}^{{p\dag}}c_{-k}^{{p\dag}}|0\rangle],
\eea

where $|\psi_{m}\rangle~(m=0,\cdots,7)$ is the eigenstate of the Hamiltonian
with corresponding eigenvalue $E_{m}$, and
$v_{j}, (j=1,\cdots,8)$ is functions of the coupling constants (see Appendix B).

\begin{figure}
\centerline{\includegraphics[width=8.5cm]{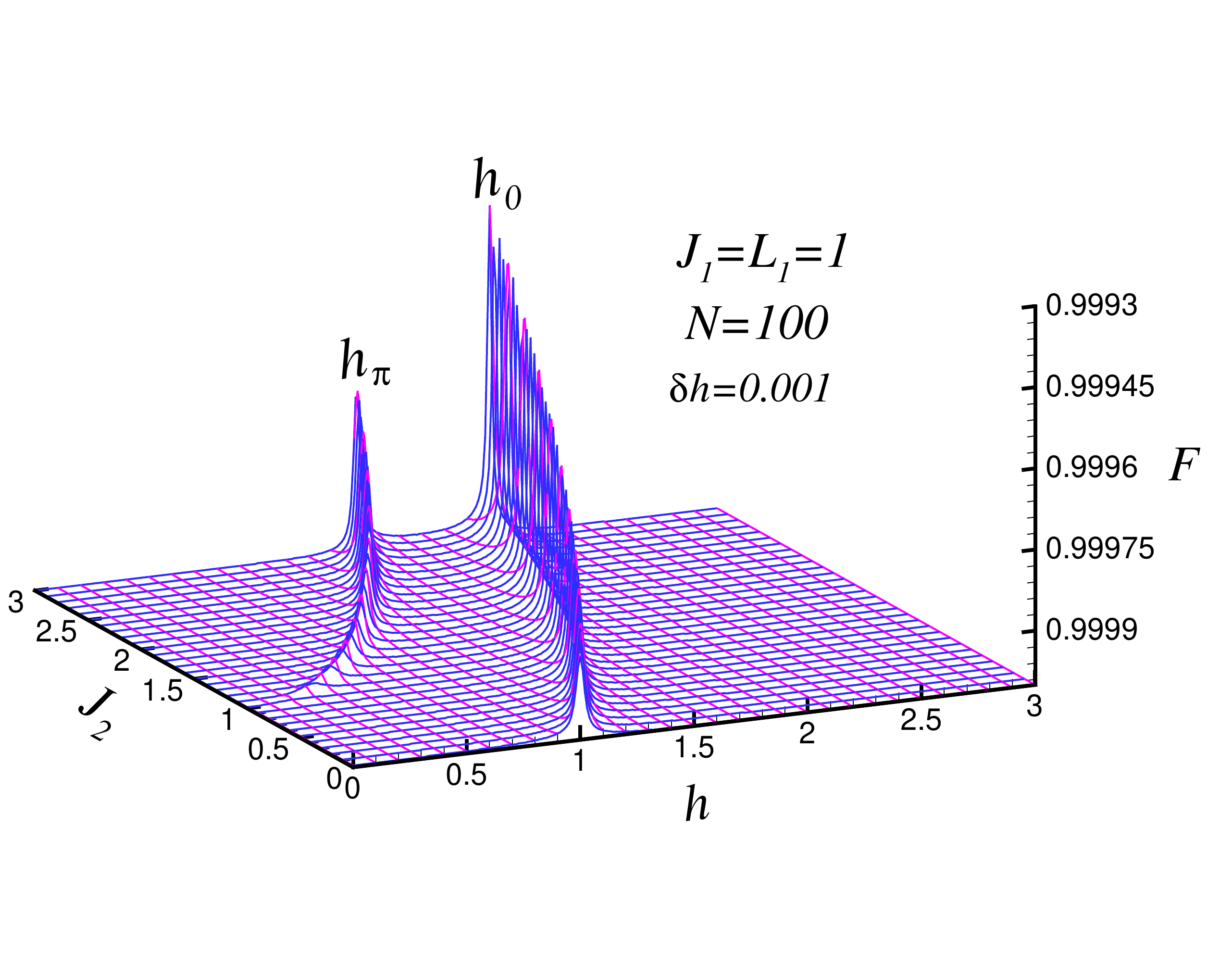}
\includegraphics[width=8.5cm]{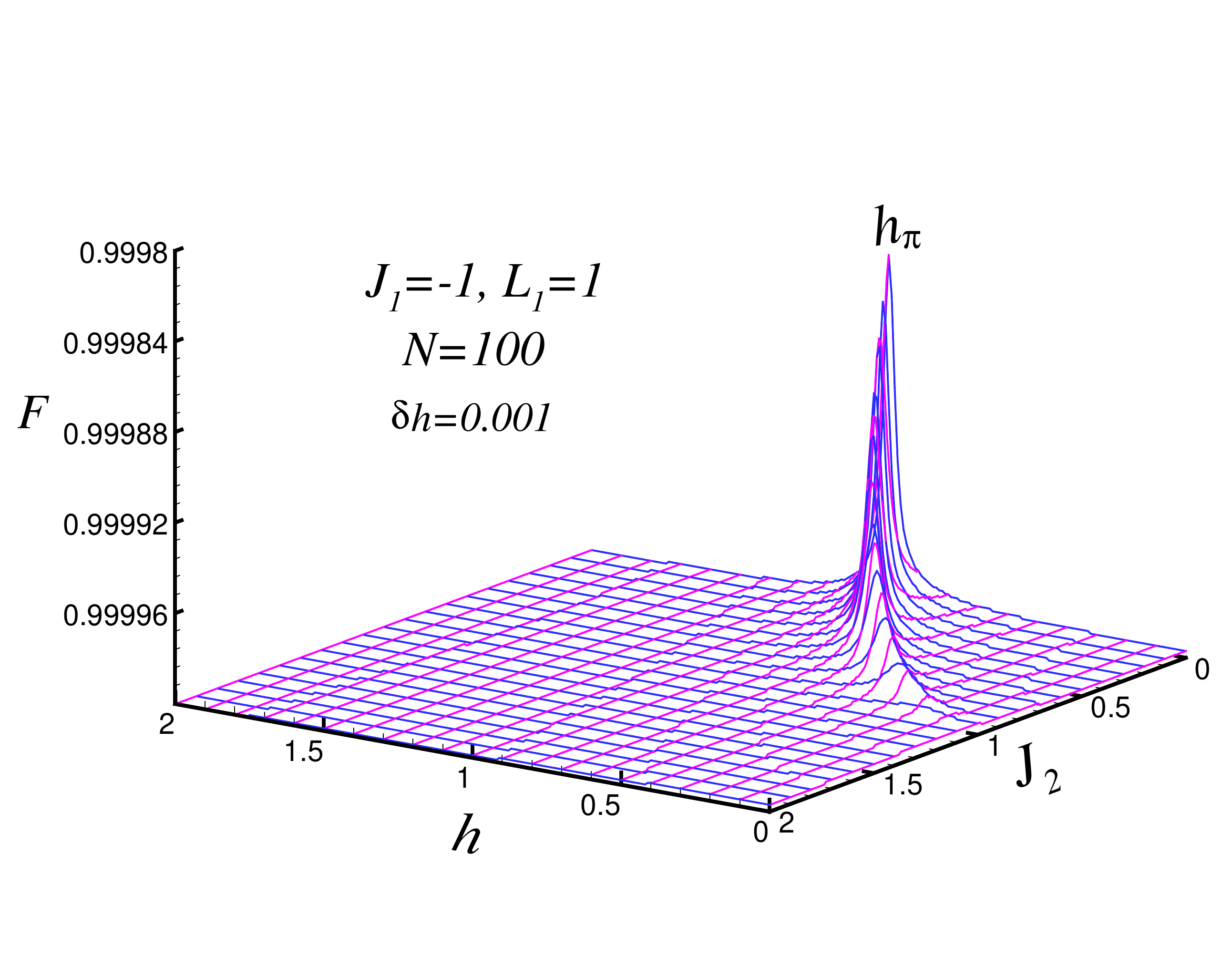}}
\caption{(Color online) Three-dimension of the ground state fidelity
of quantum compass model in a transverse field as a function of
magnetic field $h$ and Hamiltonian parameter
$J_{2}$. The parameters set as (a) $J_{1}=L_{1}=1$, $L_{2}=0$, $N=100$ and $\delta h=0.001$, (b) $J_{1}=-1$, $L_{1}=1$ $L_{2}=0$, $N=100$ and $\delta h=0.001$.}
\label{fig1}
\end{figure}

\section{Fidelity and phase diagram\label{F}}
The phase diagram can be examined by considering the fidelity and
fidelity susceptibility introduced in Ref. \cite{Zanardi}.
As mentioned in introduction, the fidelity of ground state
is defined by the overlap between the two ground state wave
functions at different parameter values and is defined as

\bea
\label{eq6}
F(h,\delta h)=|\langle \psi_{0}(h)|\psi_{0}(h+\delta h)\rangle|,
\eea

where $|\psi_{0}(h)\rangle$ is a ground state wave function of the many-body
Hamiltonian describing the system exposed to an external magnetic field $h$, while $\delta h$
is a small deviation from $h$.
The main idea is that near a QPT point there is a sharp
enhancement in the degree of distinguishability between two ground states,
corresponding to different values of the parameter space which defines
the Hamiltonian. This distinguishability can be determined by
the fidelity, which for pure states simplify to the amplitude of
inner product or overlap. The behavior can be ascribed
to a sudden change in the structure of the ground state
of the system across the quantum phase transition.
Therefore, one expects that fidelity has a minimum
at the critical point and it should contain all the information that describes
QPTs and topological order.
The drop of fidelity specify not only the position of
the critical point, but also universal information about
the transition given by the critical exponent $\nu$ which is the correlation
length exponent associated with the QCPs \cite{Zanardi}.
The response of the fidelity after an infinitesimal change of the
external parameter up to second order reads

\bea
\no
F(h,\delta h)=1-\frac{\delta h^{2}}{2}\chi_{F},
\eea

where the fidelity susceptibility ($\chi_{F}$) is defined by \cite{Venuti, You2}

\begin{figure*}
\centerline{\includegraphics[width=9cm]{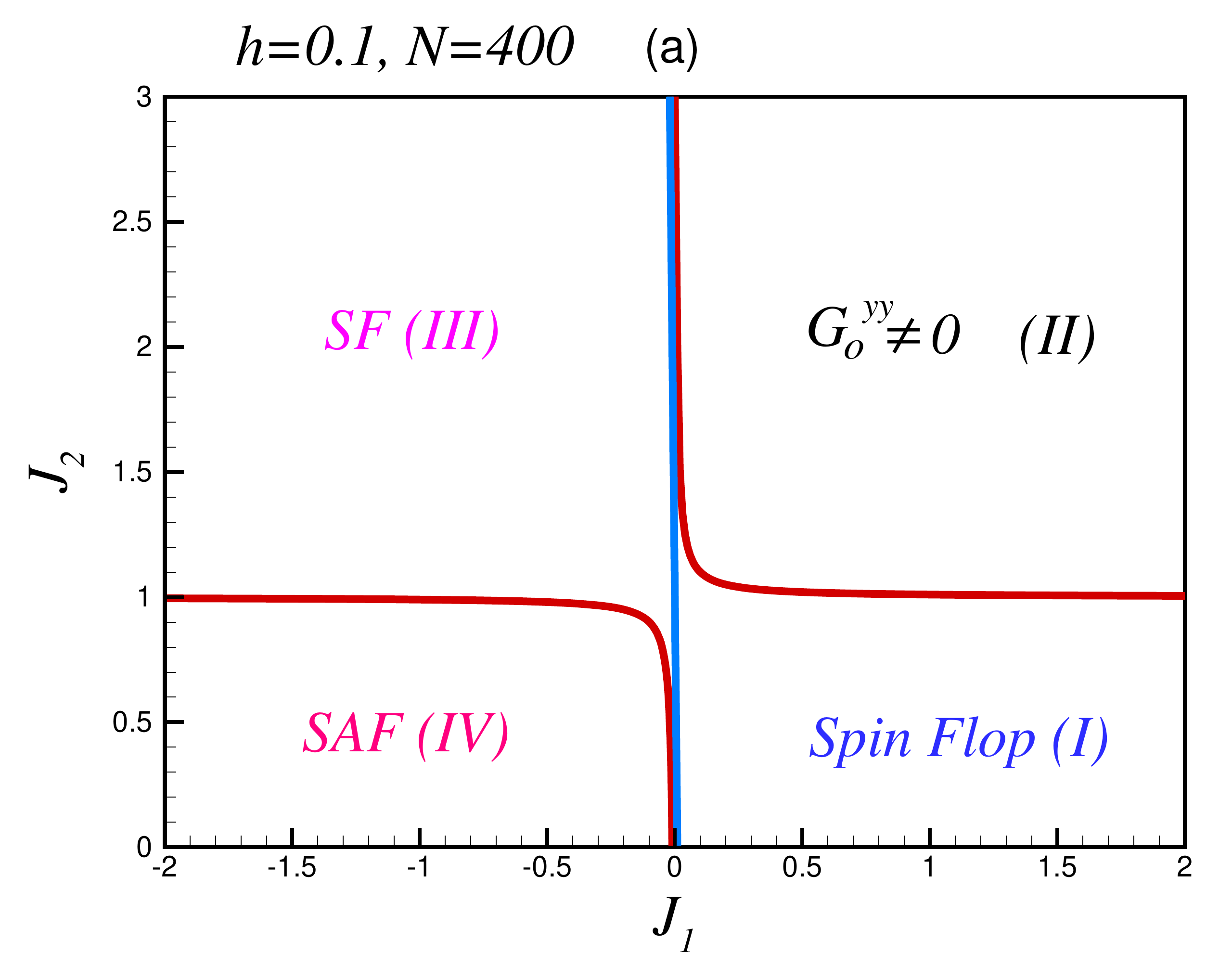}
\includegraphics[width=9cm]{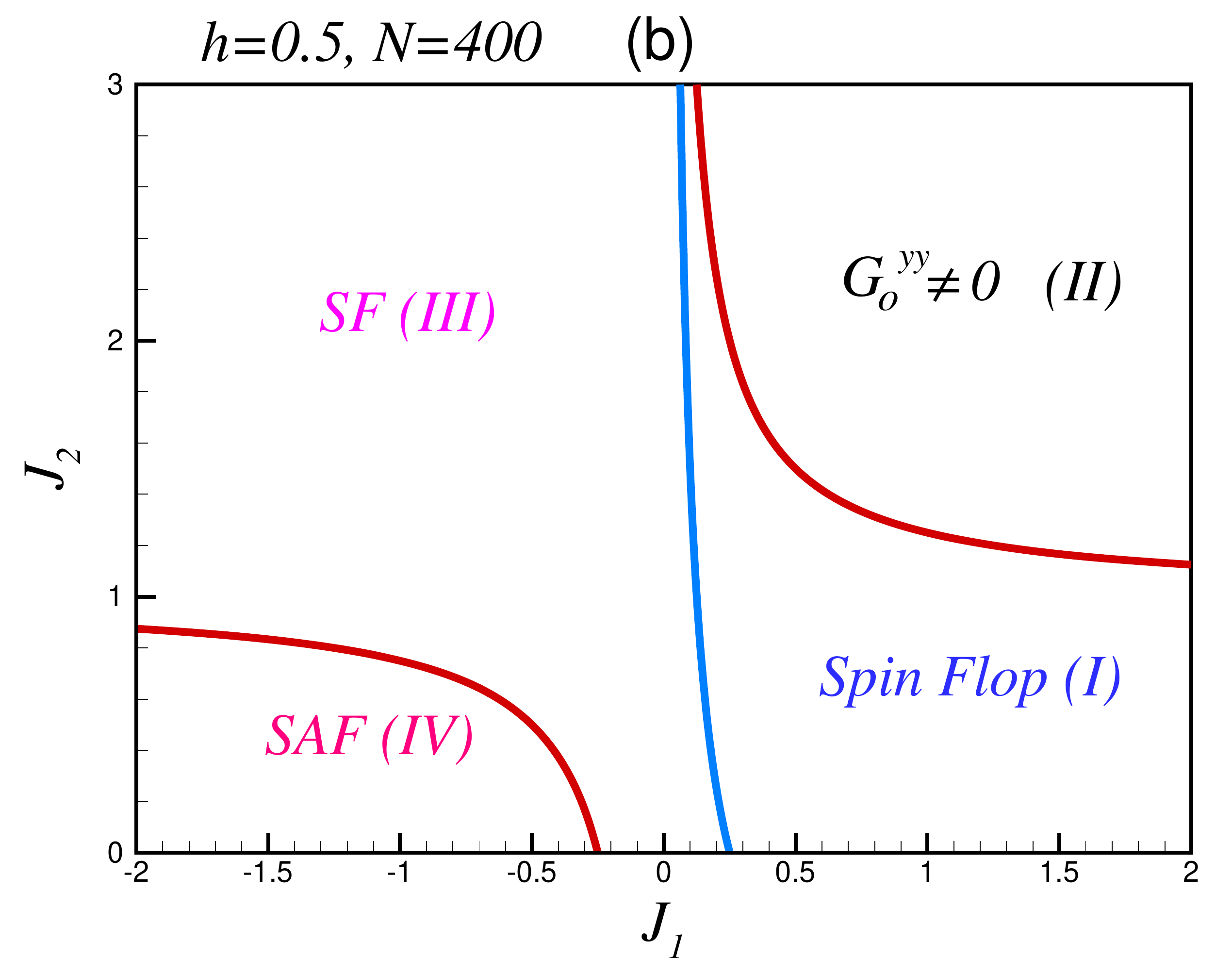}}
\caption{(Color online.) Contour plot of the fidelity of  quantum compass model
in a transverse field. The parameters set as $L_{1}=1, L_{2}=0, N=400$ and (a)
$h=0.1$, (b) $h=0.5$. The light blue line represents $h_{0}=\sqrt{J_{2}+1}$ critical
line and $h_{\pi}=\sqrt{J_{2}-1}$ critical line has been shown by the red lines.}
\label{fig2}
\end{figure*}

\bea
\label{eq7}
\chi_{F}=\langle\partial_{h}\psi_{0}(h)|\partial_{h}\psi_{0}(h)\rangle
-\langle\partial_{h}\psi_{0}(h)|\psi_{0}(h)\rangle
\langle\psi_{0}(h)|\partial_{h}\psi_{0}(h)\rangle.
\eea

If there exist more than one external parameter, this result could be generalized
to the so-called quantum geometric tensor \cite{Zanardi, Venuti, Hamma, Jafari6}.

\begin{figure*}
\centerline{\includegraphics[width=9cm]{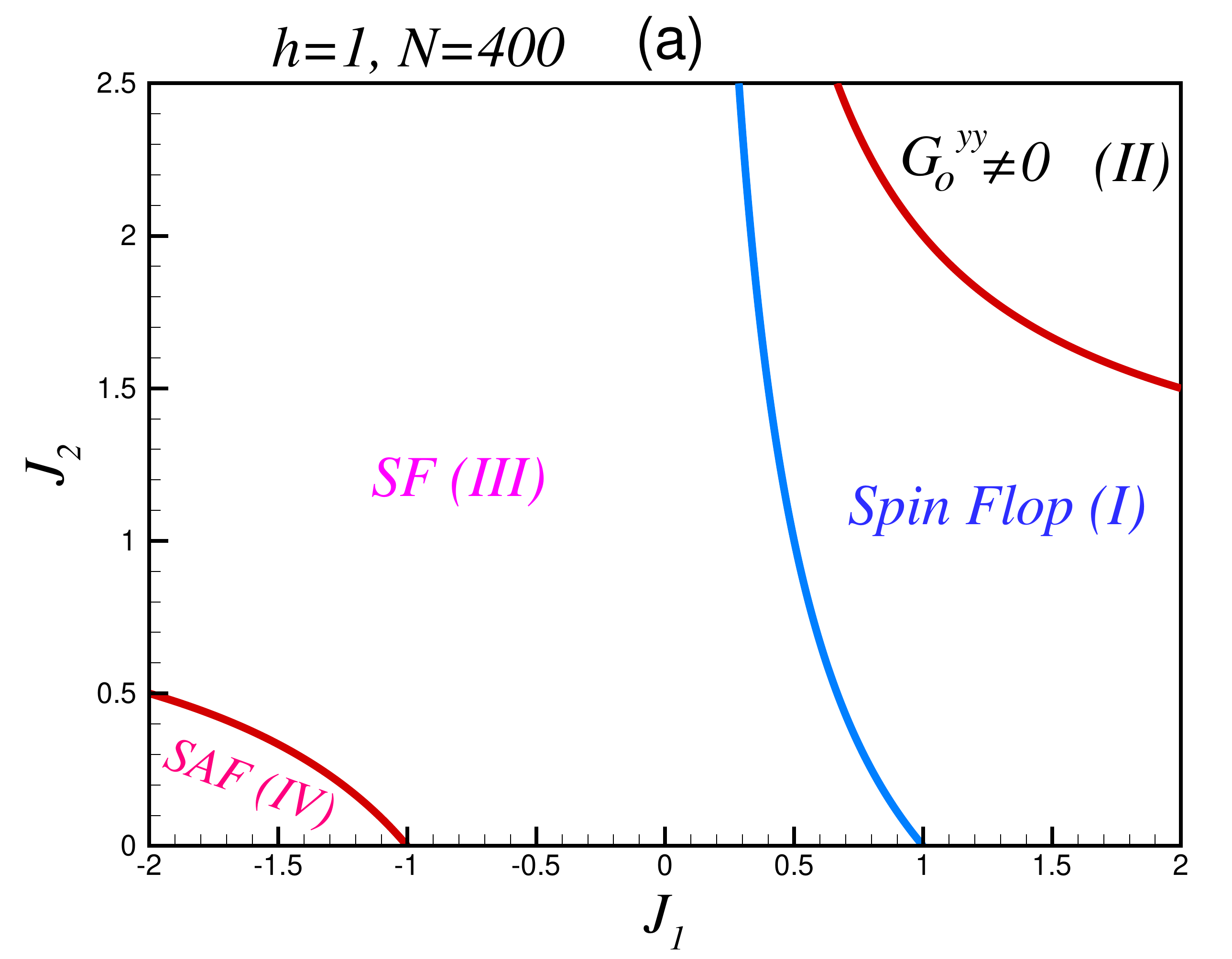}
\includegraphics[width=9cm]{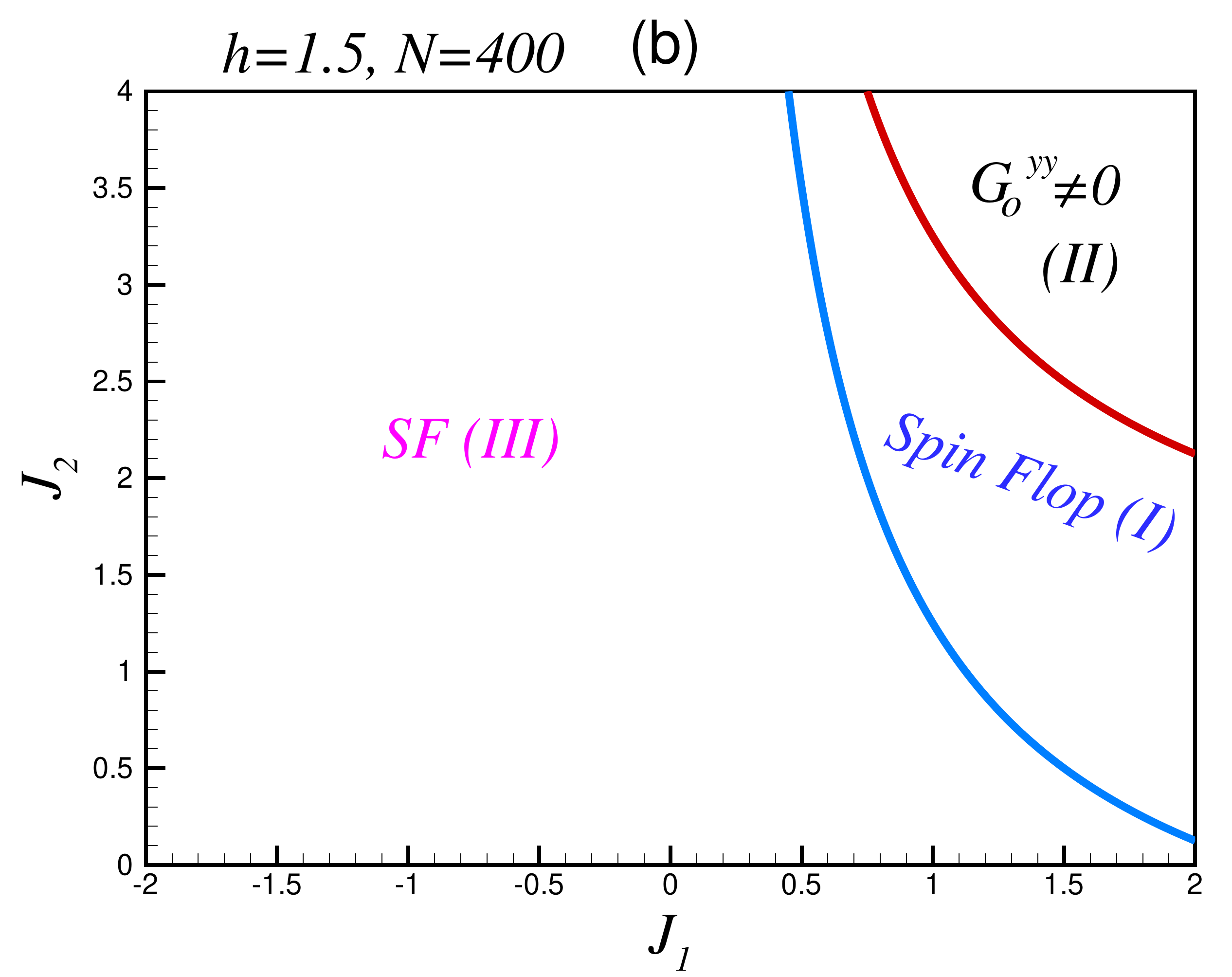}}
\caption{(Color online.) Contour plot of the fidelity of  quantum compass model
in a transverse field. The parameters set as $L_{1}=1, L_{2}=0, N=400$ and
(a) $h=1$, (b) $h=1.5$. $h_{0}=\sqrt{J_{2}+1}$ and $h_{\pi}=\sqrt{J_{2}-1}$ critical lines
have been shown by the light blue and the red lines respectively.}
\label{fig3}
\end{figure*}

The fidelity of this model has been calculated using Eqs. (\ref{eq5}) and (\ref{eq6}).
Three-dimensional panorama of the ground sate fidelity of the model has
been depicted in Figs. \ref{fig1} (a), (b) versus the magnetic field and $J_{2}$
for $J_{1}=1$ and $J_{1}=-1$, where we have set $N=100$ and $\delta h=0.001$.
Obviously, there is a sudden drop in the ground state fidelity at the QPTs lines.

In Fig. \ref{fig1} (a), one observes that the transition lines
$h_{0}=\sqrt{J_{2}+1}$ and $h_{\pi}=\sqrt{J_{2}-1}$ are characterized by two assumed lines on
the minimum parts of the fidelity surface. However, in Fig. \ref{fig1} (b) the minimum line
on the fidelity surface defined by $h_{\pi}=\sqrt{J_{2}-1}$ clearly indicates a
second order phase transition. By computing the fidelity of the one-dimensional
extended quantum compass model in a transverse filed, we find the expected critical
lines that we already discussed in the context of the exact solution and we illustrate
the phase diagram of this model in Figs. \ref{fig2} (a), (b), \ref{fig3} (a), (b).
In Ref. \cite{Jafari3} the phase diagram of this model has been investigated by
use of the gap analysis and universality of derivative of the correlation functions.
As previously mentioned, the fidelity and fidelity susceptibility of the ground state properties could
reflect different zero-temperature regions. This model is always gapful except at the critical surfaces
where the energy gap disappears. There are four gapped phases in the exchange couplings' space:

\begin{itemize}
  \item Region (I) $J_{1}>0,~0<J_{2}<J^{c}_{2}(J_{1},h)$:
  In this region for $h<h_{0}$ the ground state is in the spin-flop phase (Figs. (\ref{fig2}) and (\ref{fig3})).
  \item Region (II) $J_{1}>0,~J_{2}>J^{c}_{2}(J_{1},h)$:
  In this case there is antiparallel ordering of spin $y$ component on odd bonds ($G_{o}^{yy}\neq0$)\cite{Eriksson,Mahdavifar,Jafari3} for $h<h_{\pi}$.
  In this region tuning $J_{2}$ forces the system go into a spin-flop phase (region (I)).
  \item Region (III) $J_{1}<|J^{c}_{1}(J_{2},h)|$,~$J_{2}>J^{c}_{2}(J_{1},h)$:
  In this region the ground state is the ferromagnetically polarized state (Saturate Ferromagnetic)
  along the magnetic field (SF).
  (Figs. (\ref{fig2}) and (\ref{fig3})).
  \item Region (IV) $J_{1}<0,~J_{2}<J^{c}_{2}(J_{1},h)$:
  In this region the ground state is in the strip antiferromagnetic (SAF) phase for $h<h_{\pi}$.
 \end{itemize}

\begin{figure*}
\centerline{\includegraphics[width=9cm]{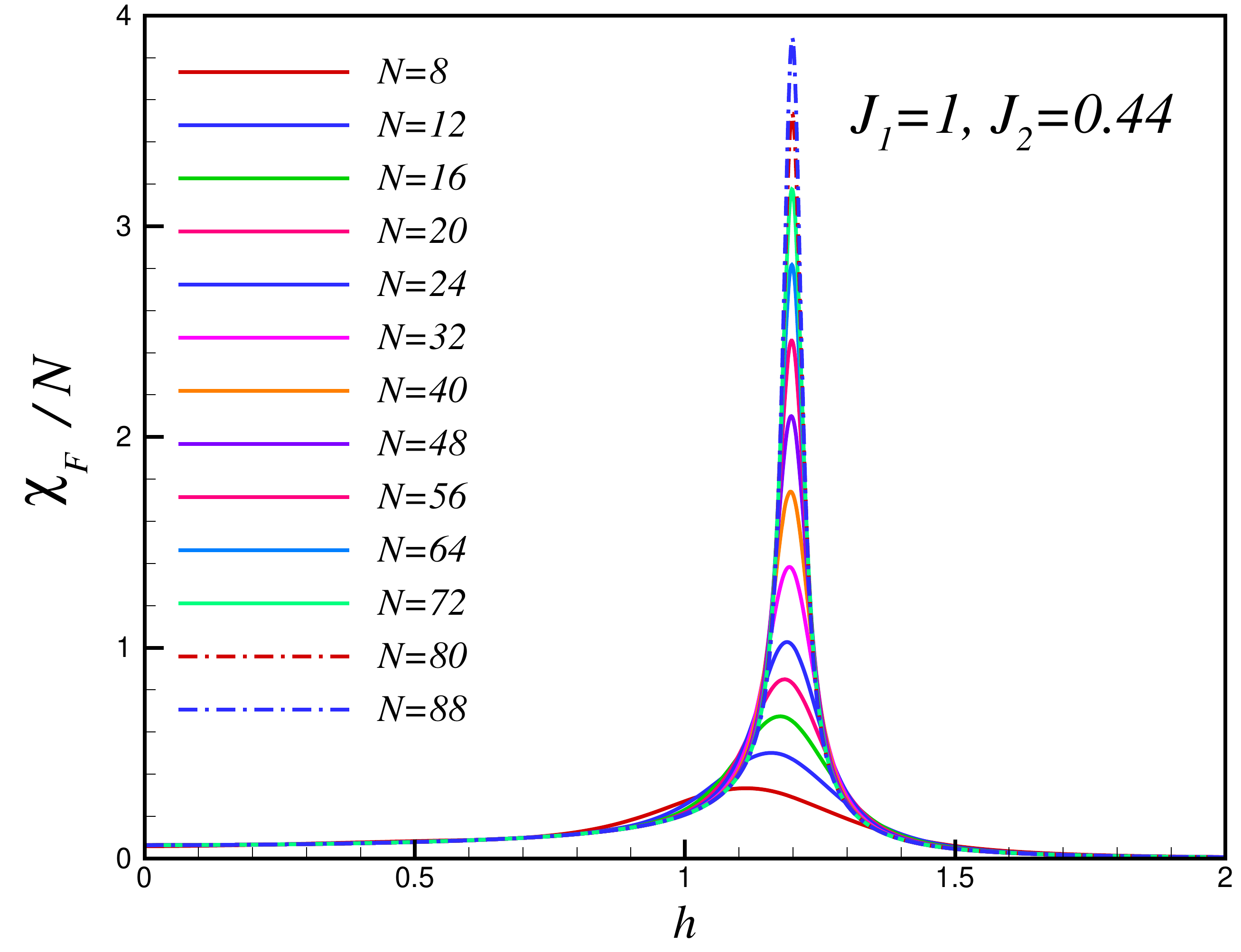}
\includegraphics[width=9cm]{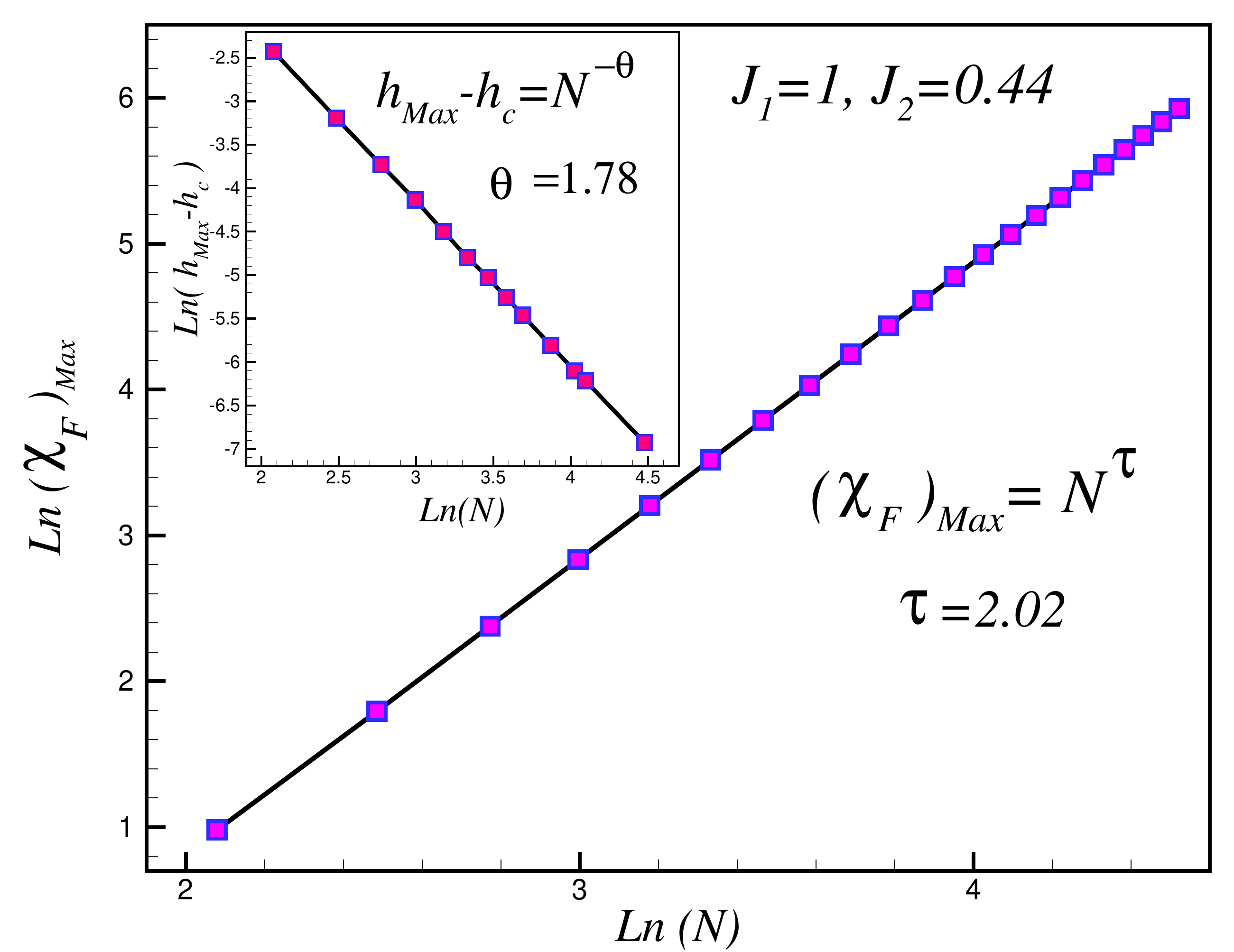}}
\caption{(Color online) (a) Fidelity susceptibility per particle of quantum
compass model in a transverse field as a functions of h for various
system size for $J_{1}=1, J_{2}=0.44$. (b) Scaling of the maximum of $\chi_{F}$
in terms of system size ($N$). Inset: Scaling of the position
($h_{Max}$) of $\chi_{F}$ for different-length chains  where
$h_{Max}$ is the position of maximum in Fig.\ref{fig4} (a).}
\label{fig4}
\end{figure*}

\begin{figure*}
\centerline{\includegraphics[width=9cm]{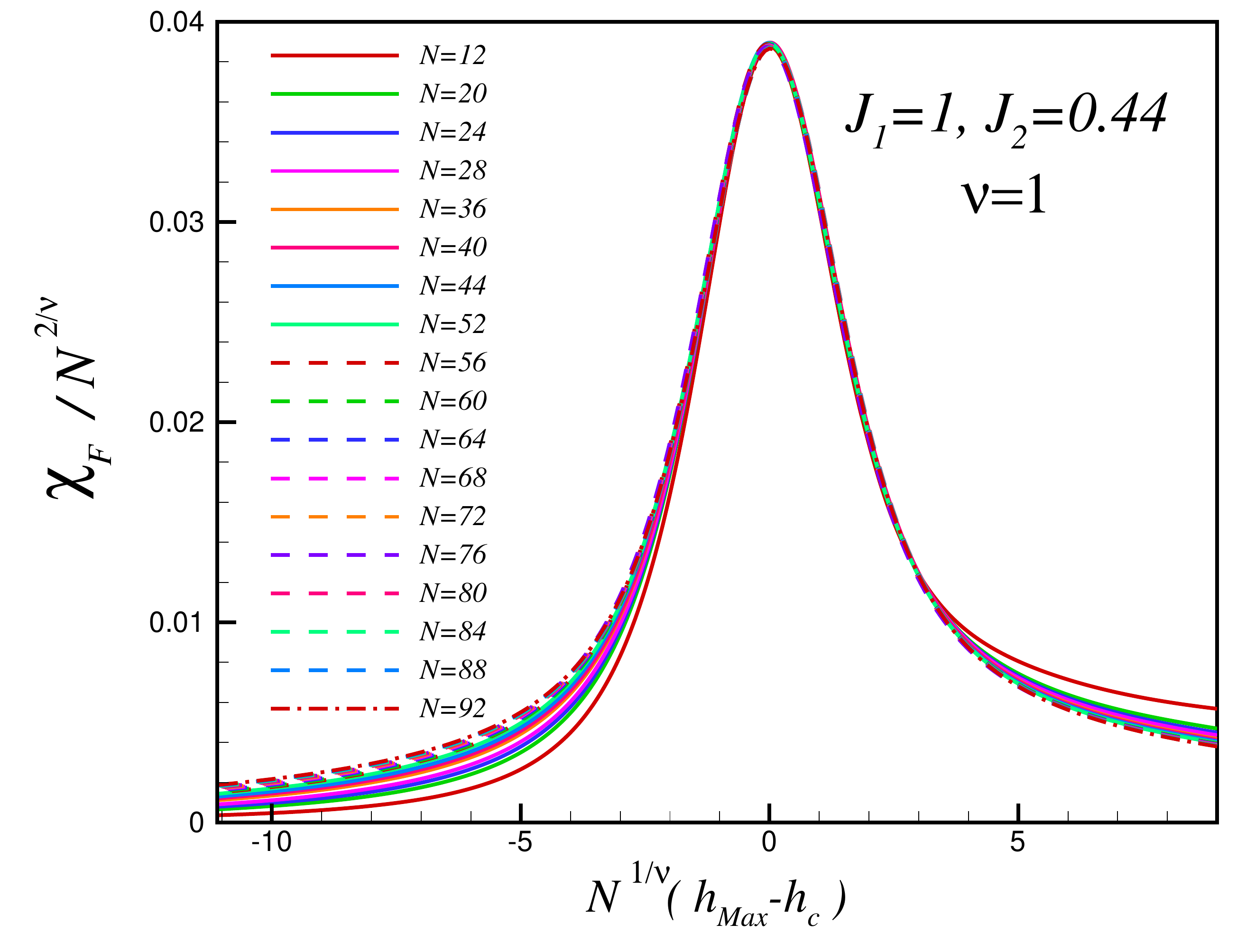}
\includegraphics[width=9cm]{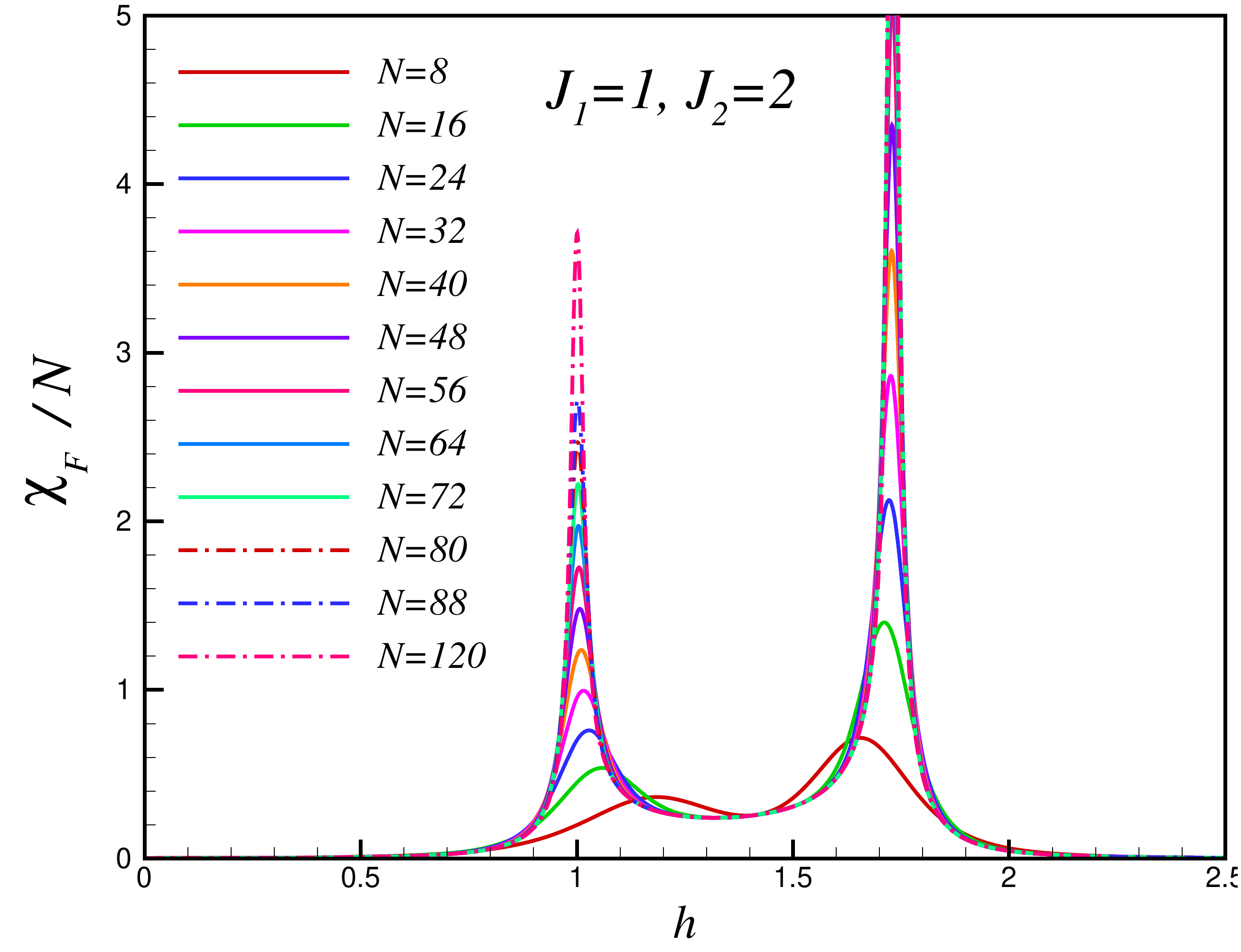}}
\caption{(Color online) (a) The finite-size scaling analysis for the
case of logarithmic divergence around the maximum point
($h_{Max}$) for $J_{1}=1, J_{2}=0.44$. The fidelity susceptibility,
considered as a function of system size and couplings,
collapses on a single curve for different lattice sizes. (b) Fidelity susceptibility of the model as a
function of h for different system sizes for $J_{1}=1, J_{2}=2$.}
\label{fig5}
\end{figure*}

\begin{table}
\begin{center}
\label{table1}
\caption{The critical exponents $\theta$, $\tau$,  $\nu$. The scaling behavior of
the susceptibility in the vicinity of the critical points.}
\begin{tabular}
{|ccccc|}
\hline
$Exchange~couplings$  & ~~$Critical~points$  & ~$\theta$ & ~$\tau$ &  ~$\nu$\\
\hline
\hline
$J_{1}=1, J_{2}=0.44$ & ~~$h_{c}=h_{0}=1.2$  & ~$1.78$ & ~$2.02$   & ~$1.00$  \\
\hline
$J_{1}=1, J_{2}=2$    & ~~$h_{c_{1}}=h_{\pi}=1.0$ & ~$1.90$    & ~$1.97$     & ~$1.00$  \\
\hline
$J_{1}=1, J_{2}=2$    & ~~$h_{c_{2}}=h_{0}=\sqrt{3}$ & ~$1.90$ & ~$2.01$   & ~$0.985$  \\
\hline
$J_{1}=-1, J_{2}=0.51$  & ~~ $h_{c}=h_{\pi}=0.7$ & ~$2.70$ & ~$1.93$   & ~$1.01$  \\
\hline
\end{tabular}
\end{center}
\end{table}

\section{Universality and scaling of fidelity susceptibility\label{USFS}}

It is expected that the fidelity susceptibility ($\chi_{F}$) probes the QPTs.
Then it will be helpful to study the universality and scaling behavior
of $\chi_{F}$ to better understand the properties of the fidelity,
and the relation between fidelity and quantum criticality.
In this section we investigate the scaling behavior of the fidelity susceptibility
by the finite size scaling approach.
In Fig. (\ref{fig4}) (a) the two dimensional plot of the fidelity susceptibility per
particle ($\chi_{F}/N$) has been shown versus the magnetic field for different
system sizes for $J_{1}=1, J_{2}=0.44$.

Although there is no real divergence for finite lattice
size, but the curves exhibit marked anomalies with height of peak increasing
with the system size and in thermodynamics limit $\chi_{F}/N$ diverges as the critical
point is touched. More information can be obtained when the maximum values of each
plot and their positions are examined.
As it manifests the divergences of $\chi_{F}$ occurs at $h_{c}=1.2$
where exactly correspond to the critical point that has been obtained using the energy
gap analysis ($h_{c}=h_{0}=\sqrt{1.44}$). Our investigation manifest the scaling
behavior of $\chi_{F}$ at the maximum point versus $N$. In this way we have plotted
the scaling behavior of $\chi_{F}|_{h_{Max}}$ in  Fig. \ref{fig4} (b), which shows a linear
behavior of $\ln(\chi_{F}|_{h_{Max}})$ versus $\ln(N)$ with the exponent $\tau\simeq2.02\pm0.01$.

A more detailed analysis shows that the position of the maximum point ($h_{Max}$) of
$\chi_{F}$ tends toward the critical point like $h_{Max}=h_{c}+N^{-\theta}$
($\theta=1.78\pm0.01$) which has been plotted in the inset of Fig. \ref{fig4} (b).

To study the scaling behavior of fidelity susceptibility around the critical
points, we perform finite-scaling analysis, since the maximum value of $\chi_{F}$
scales logarithmically.
Then, by choosing a proper scaling function and taking into account the distance of the
maximum of $\chi_{F}$ from the critical point, it is possible to make all the data for
the value of $\chi_{F}/N^{2/\nu}$ as a function of $N^{1/\nu}(h_{Max}-h_{c})$ for different
$N$ collapse onto a single curve.
The analysis of the finite-size scaling is shown in Fig. \ref{fig5} (a) for several typical lattice sizes.
It is clear that the different curves which correspond to various system sizes collapse
to a single universal curve as expected from the finite size scaling ansatz.
Our result shows that $\nu=1$ exactly corresponds to the correlation length exponent
of Ising model in a transverse field ($\nu=1$).

\begin{figure*}
\includegraphics[width=9cm]{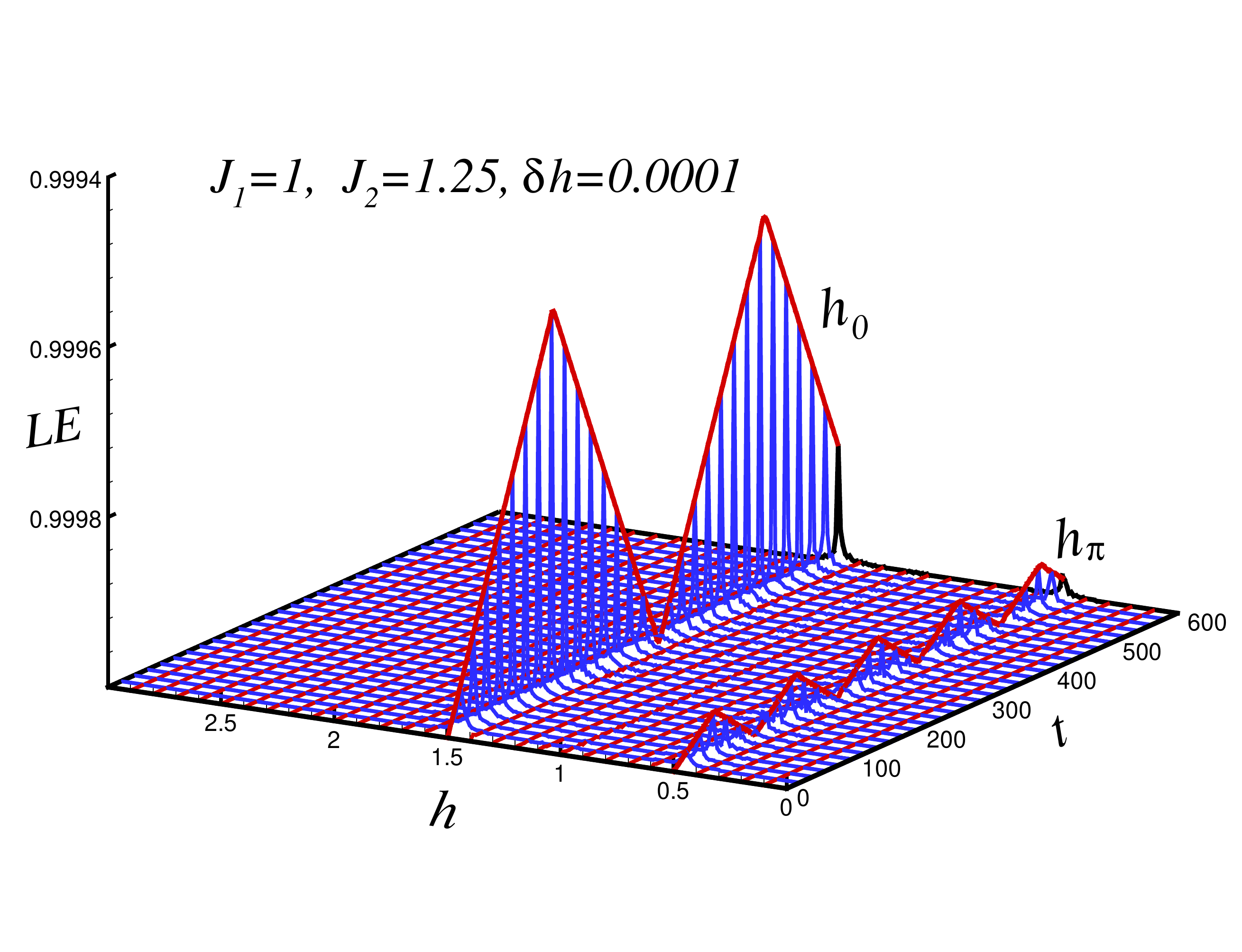}
\caption{(Color online) Three-dimension of the Loschmidt echo
as a functions of magnetic field and time, where the Hamiltonian
parameters set as $J_{1}=1, J_{2}=1.25, L_{1}=1, L_{2}=0$ for $N=400$ and
$\delta h=0.0001$.}
\label{fig6}
\end{figure*}

\begin{figure*}
\centerline{\includegraphics[width=6.2cm]{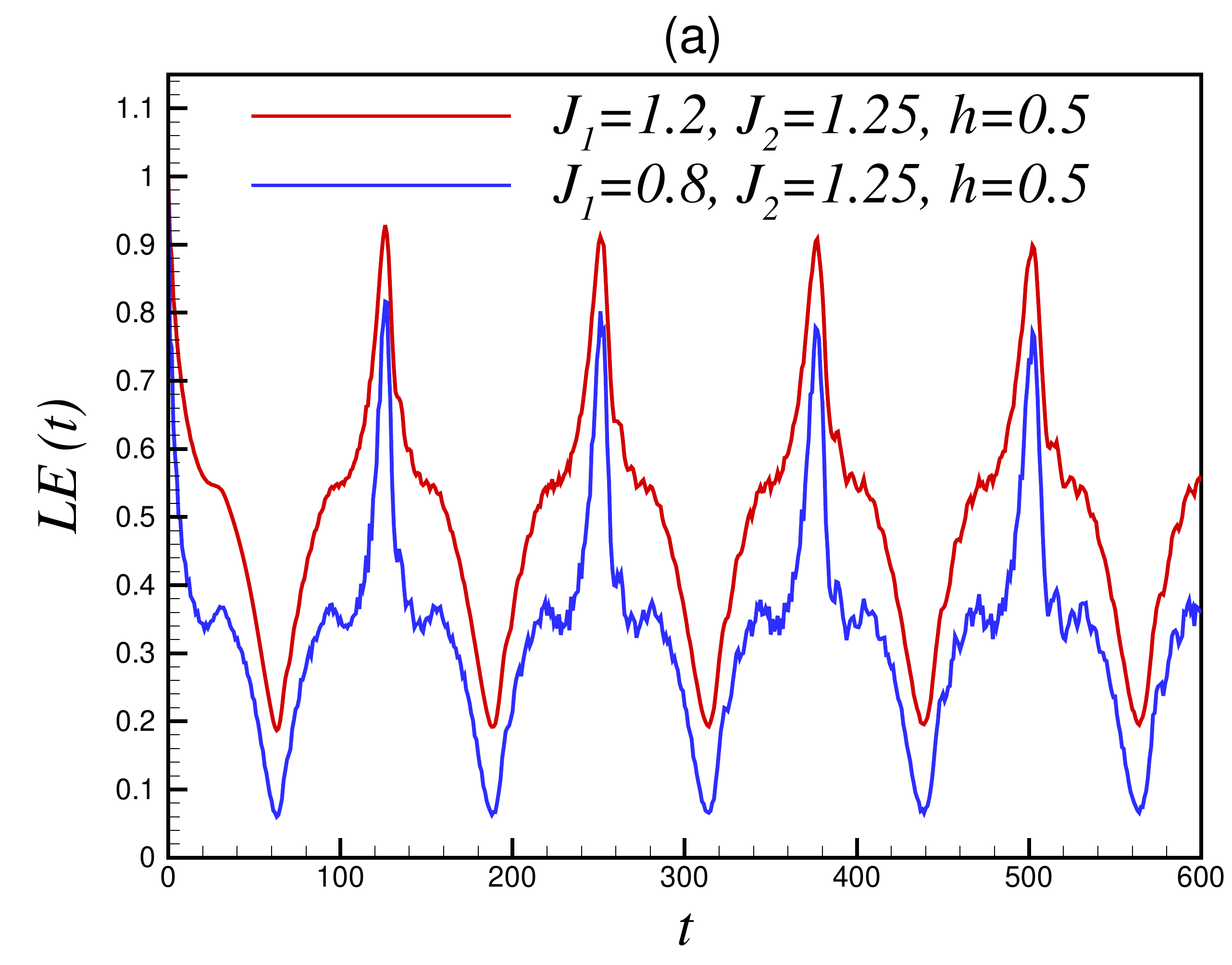}
\includegraphics[width=6.2cm]{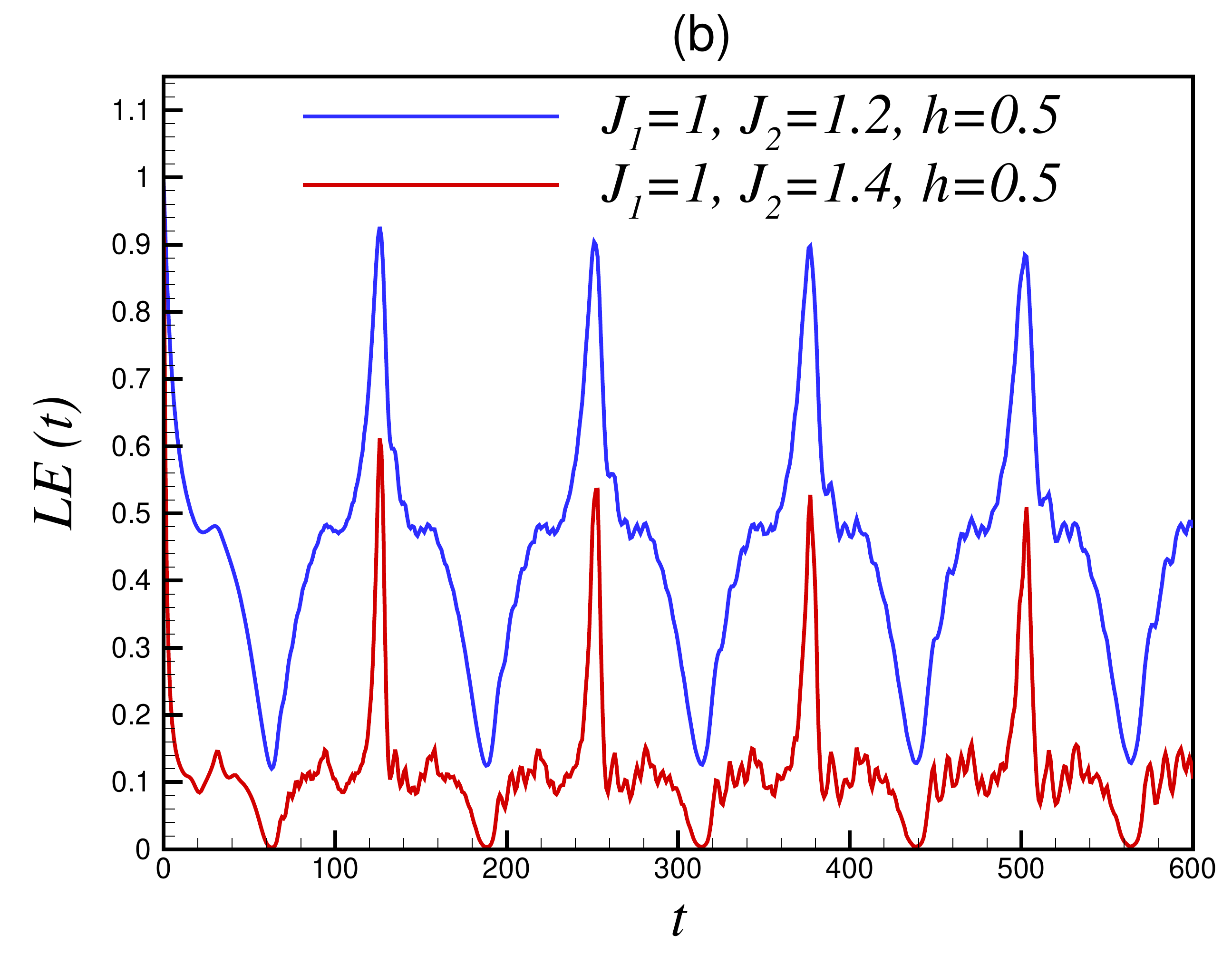}
\includegraphics[width=6.2cm]{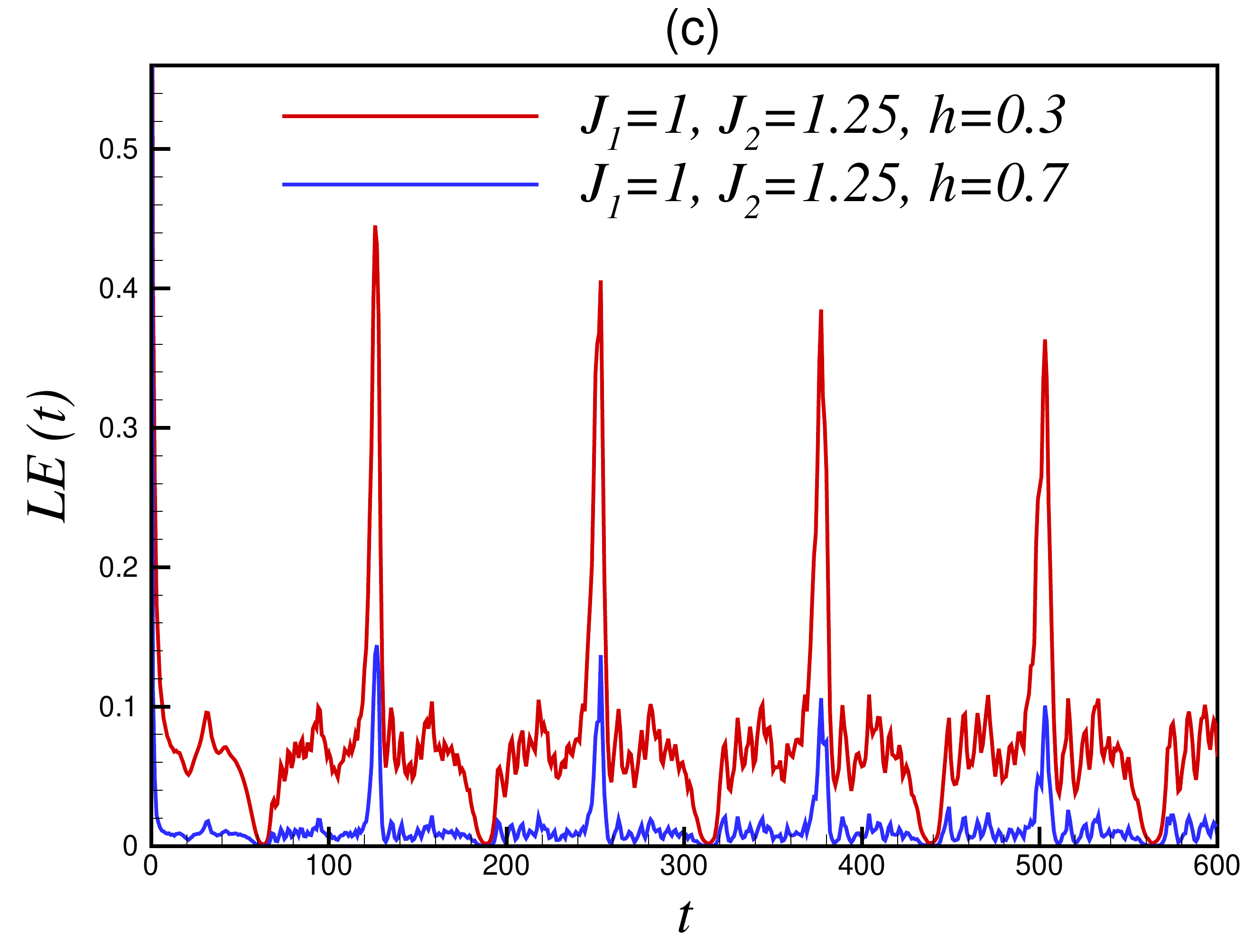}}
\caption{(Color online.) Variation of ground state LE versus the
time $t$ to the critical point $J_{1}=1, J_{2}=1.25, L_{1}=1, L_{2}=0, h=0.5$ with $N=400$,
starting from different values of the coupling constants and
magnetic field (a) $J_{1}=0.8$, $J_{1}=1.2$, (b) $J_{2}=1.2$,
$J_{2}=1.4$, (c) $h=0.3$ and $h=0.7$.}
\label{fig7}
\end{figure*}

\begin{figure*}
\centerline{\includegraphics[width=6.2cm]{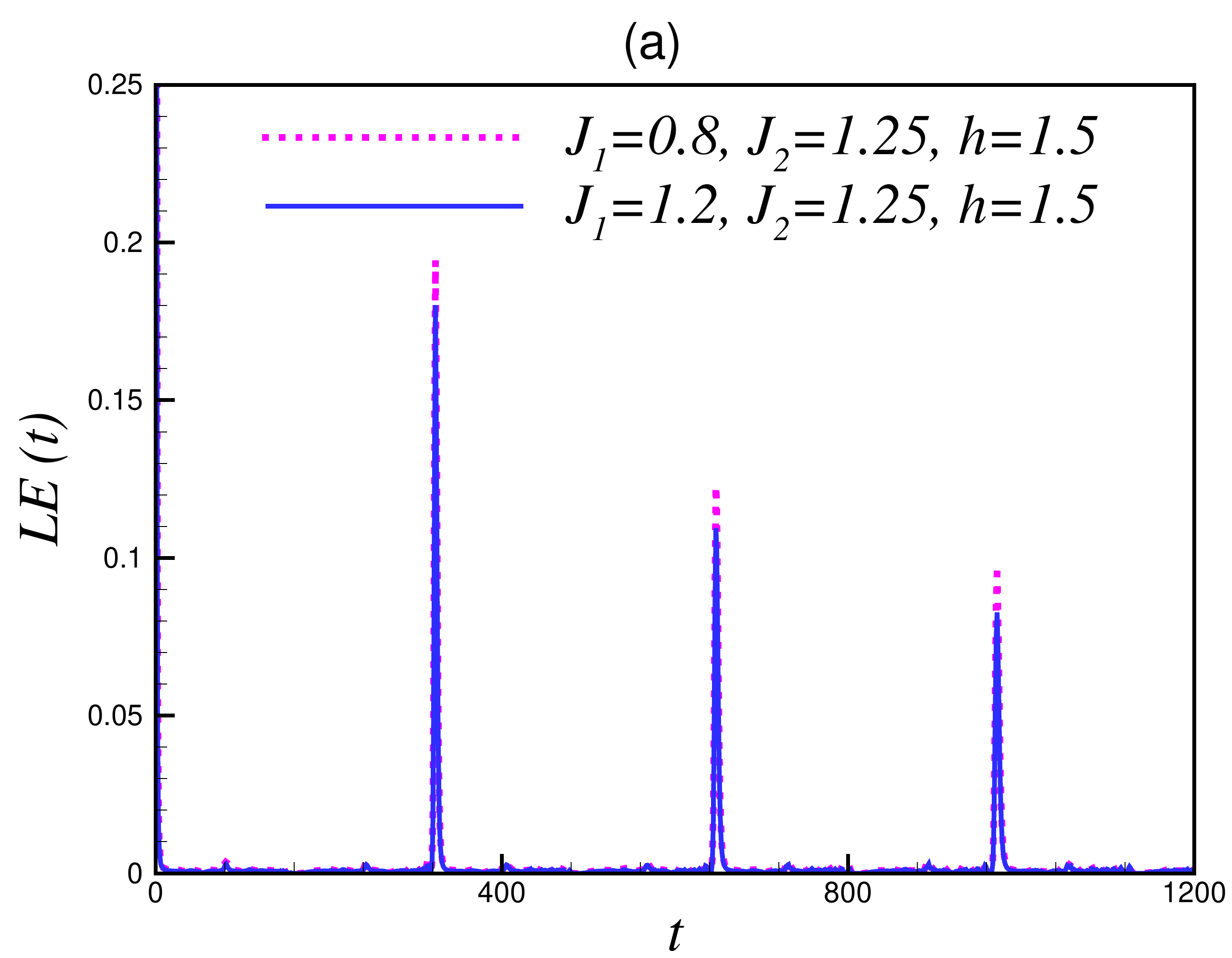}
\includegraphics[width=6.2cm]{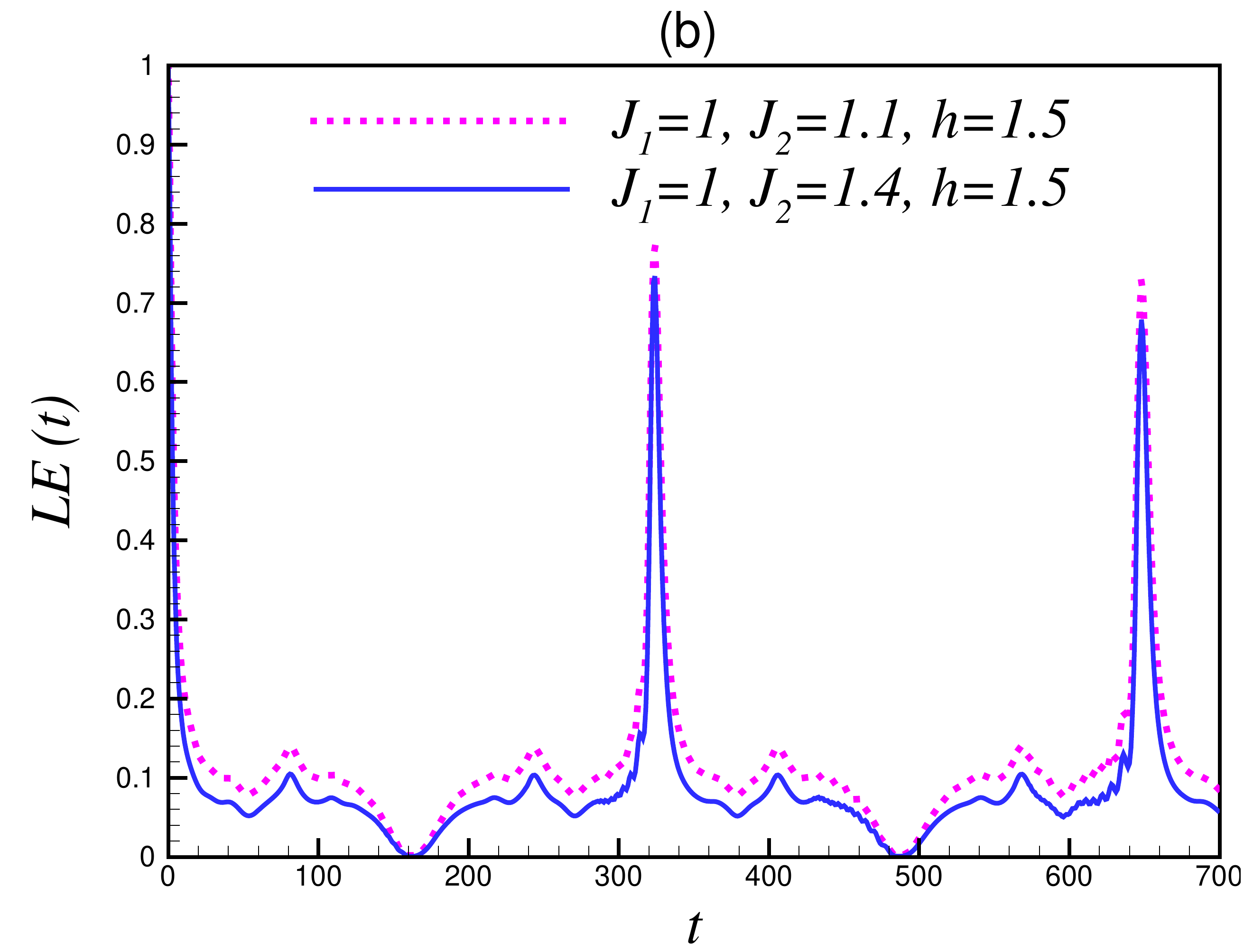}
\includegraphics[width=6.2cm]{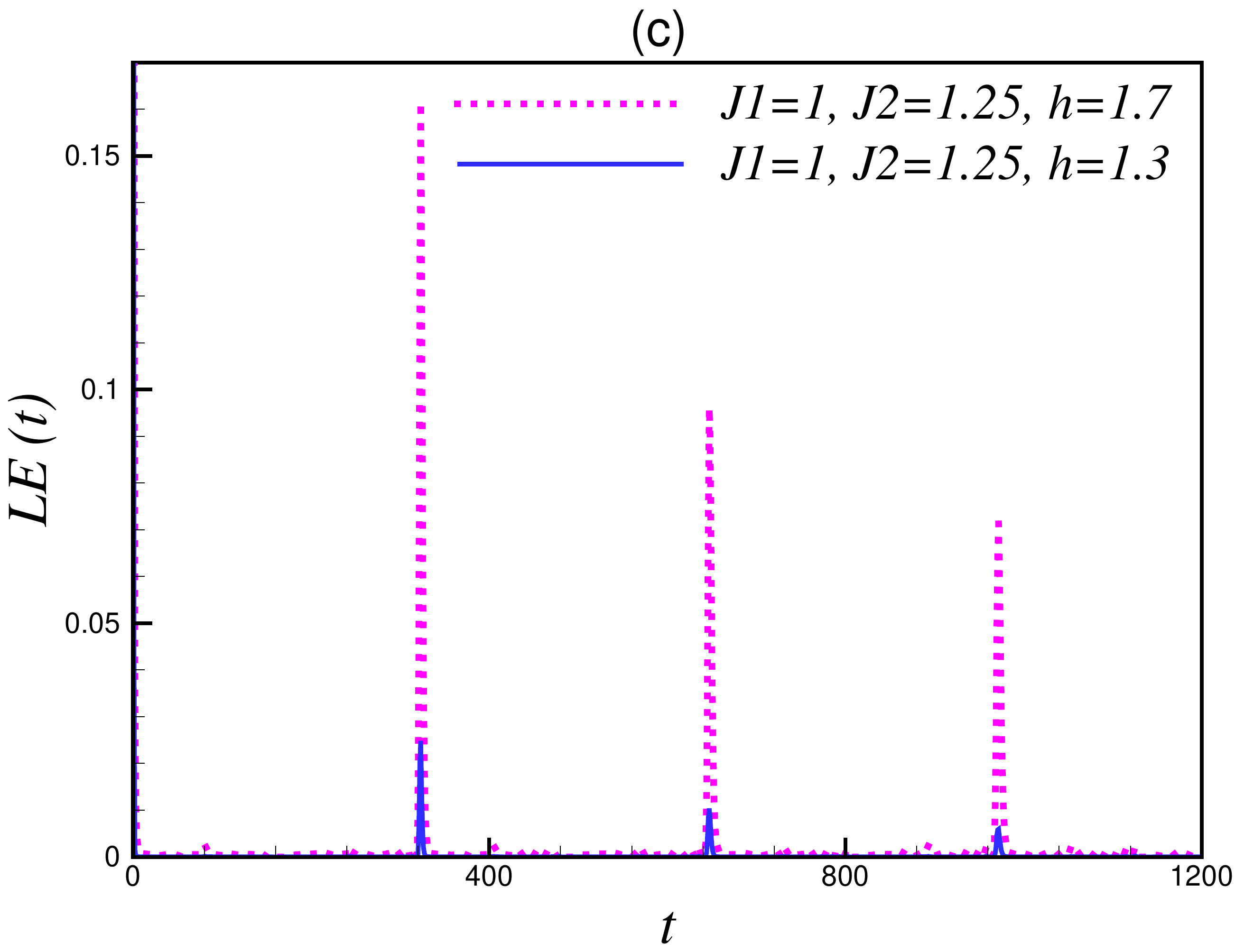}}
\caption{(Color online.) The LE for the quenched quantum compass model
to the critical point $J_{1}=1, J_{2}=1.25, L_{1}=1, L_{2}=0, h=1.5$ with $N=400$, starting
from different values of parameters (a) $J_{1}=0.8$ and $J_{1}=1.2$, (b) $J_{2}=1.1$
and $J_{2}=1.4$, (c) $h=1.3$ and $h=1.7$.}
\label{fig8}
\end{figure*}

\begin{figure*}
\centerline{\includegraphics[width=6.2cm]{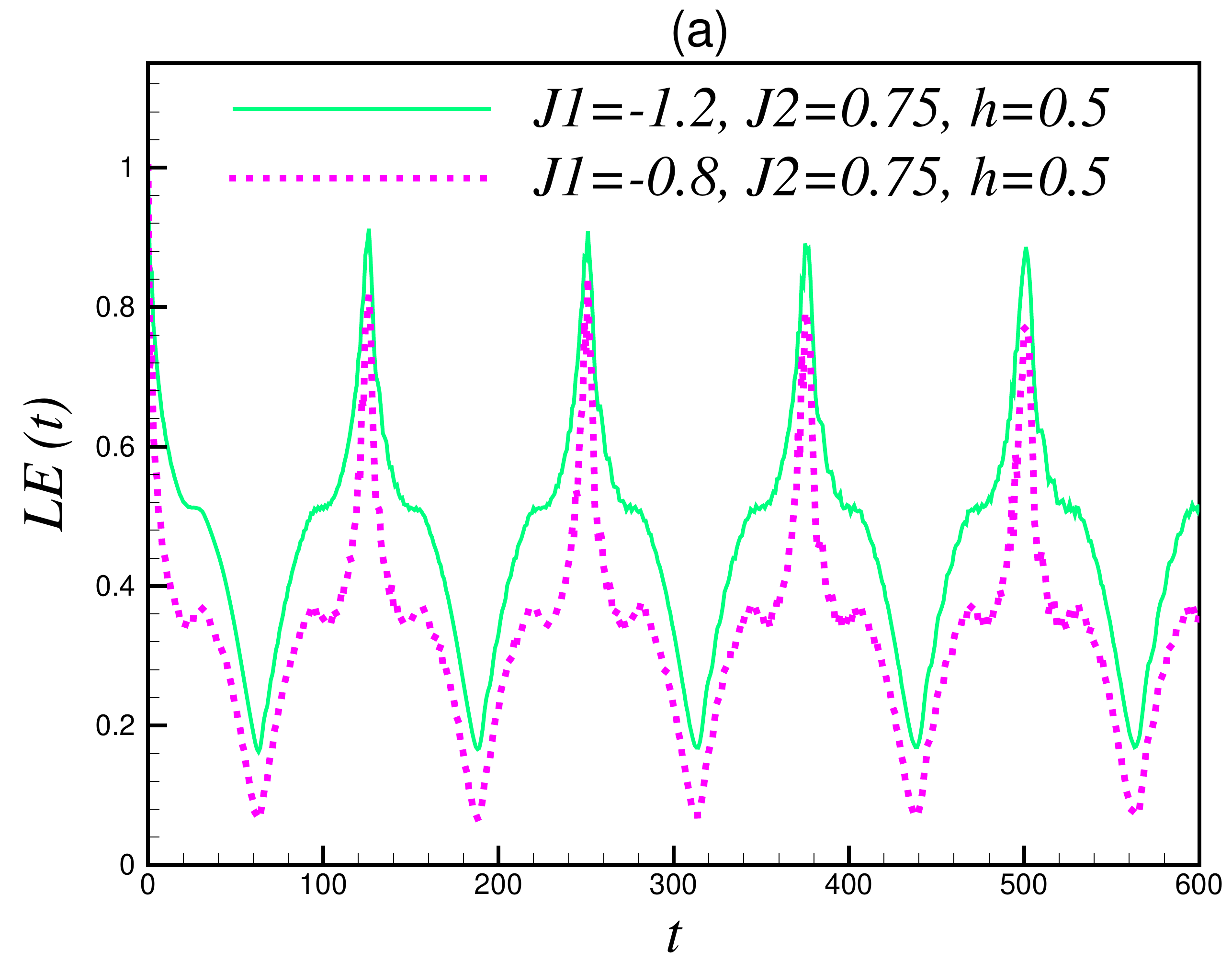}
\includegraphics[width=6.2cm]{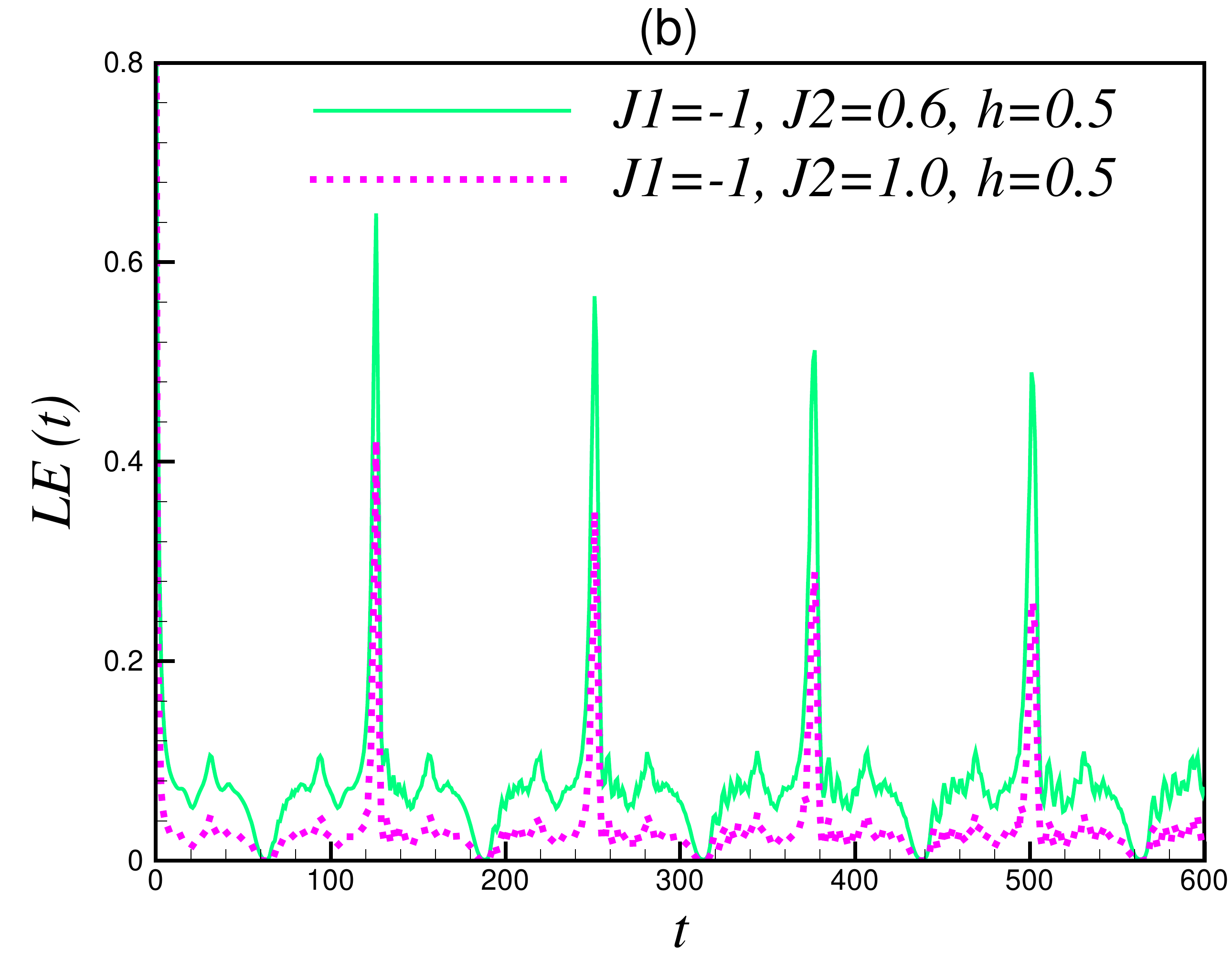}
\includegraphics[width=6.2cm]{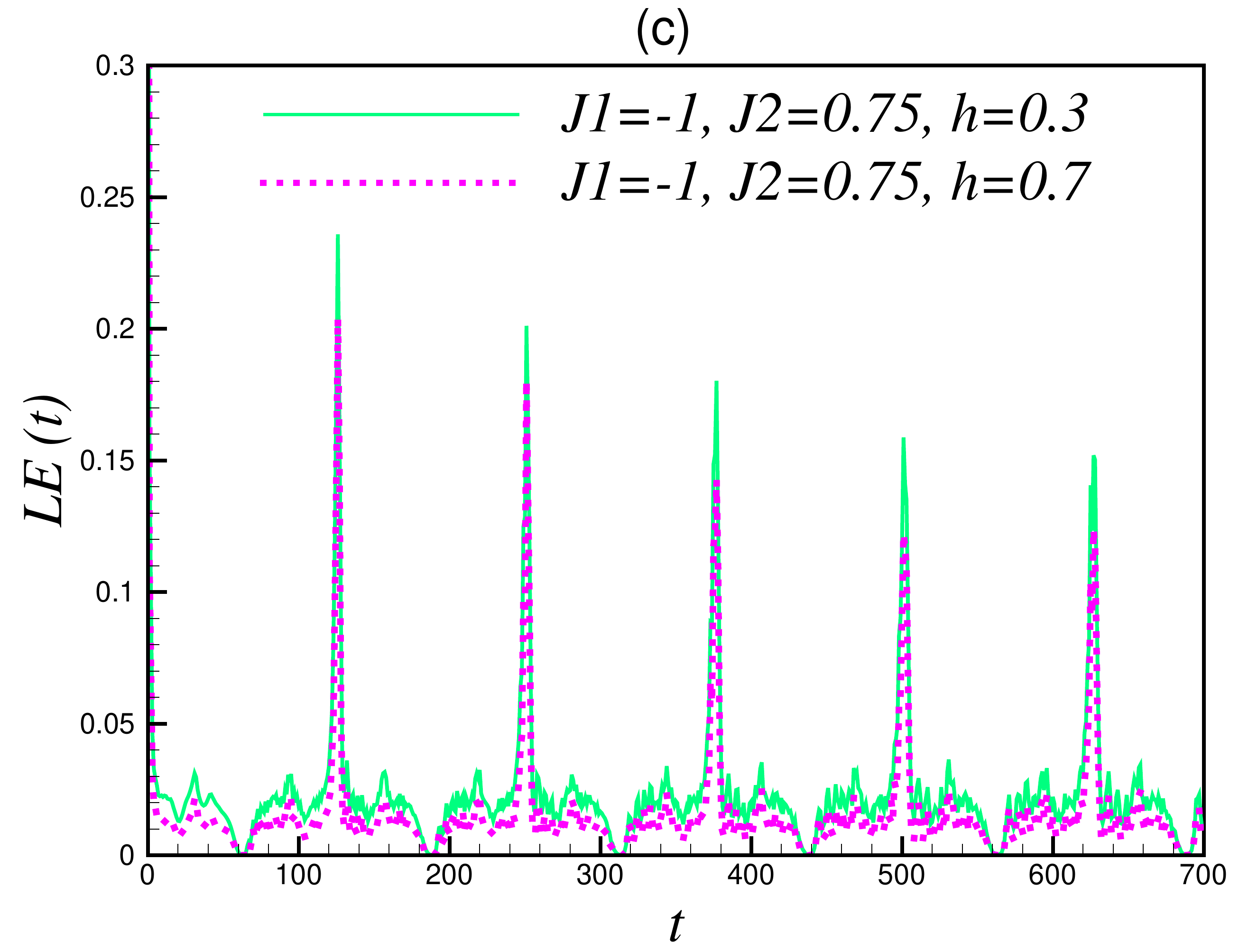}}
\caption{(Color online.) The LE starting from the SAF and SF phases respectively quenched to
the critical point $J_{1}=-1, J_{2}=0.75, L_{1}=1, L_{2}=0, h=0.5$ with $N=400$,
(a) $J_{1}=-1.2$ and $J_{1}=-0.8$, (b) $J_{2}=0.6$ and $J_{2}=1$, (c) $h=0.3$ and $h=0.7$.}
\label{fig9}
\end{figure*}

\begin{figure*}
\centerline{\includegraphics[width=6.2cm]{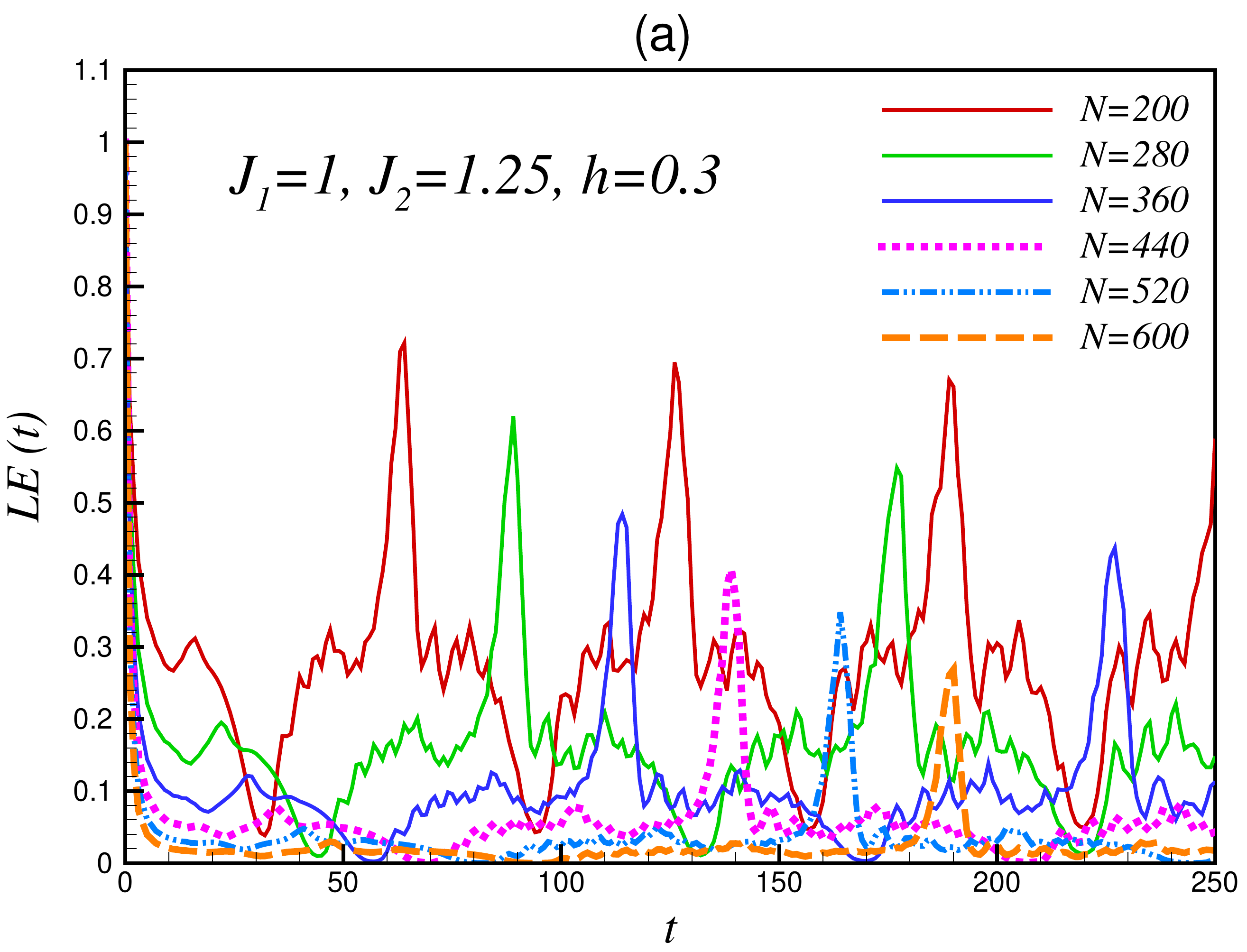}
\includegraphics[width=6.2cm]{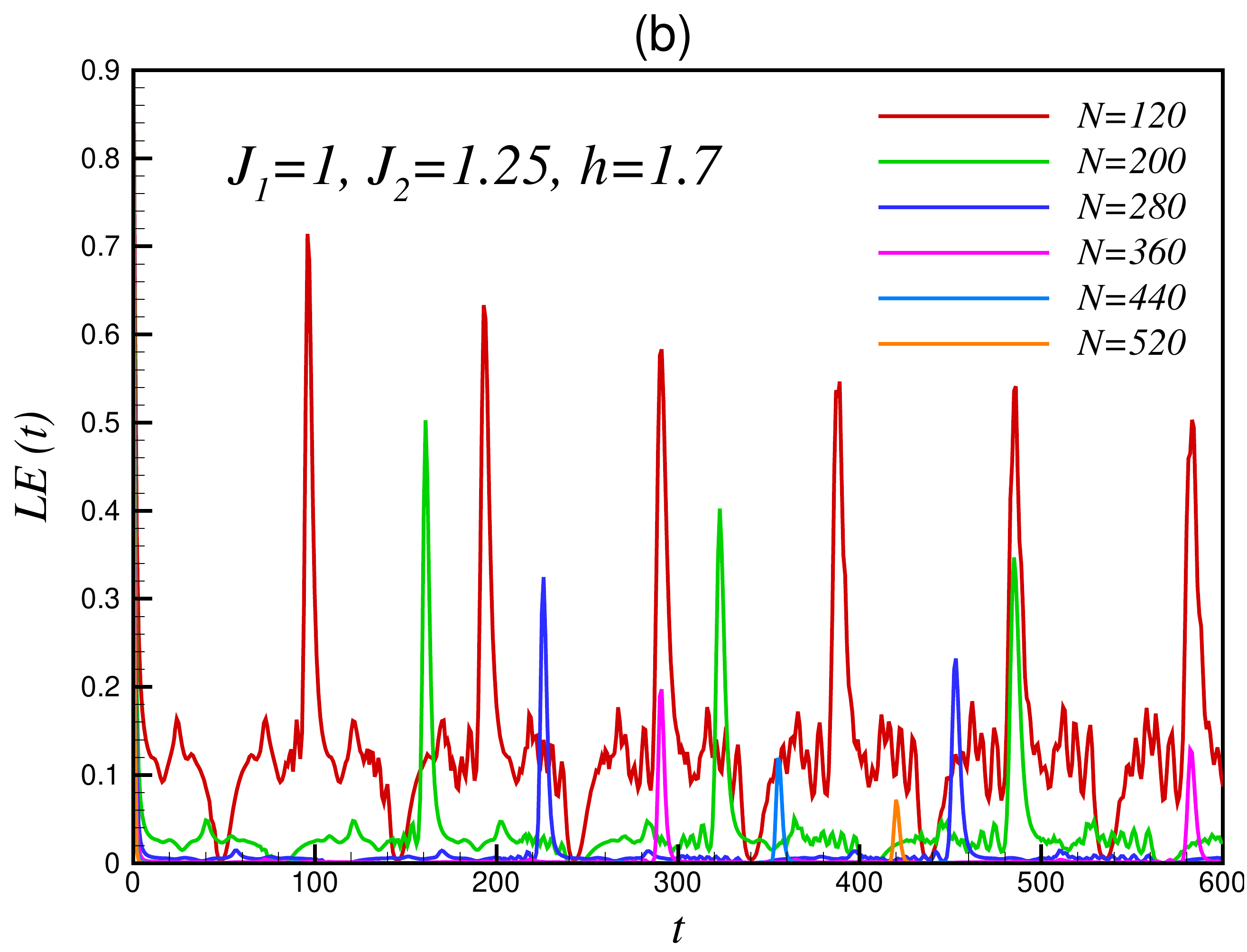}
\includegraphics[width=6.2cm]{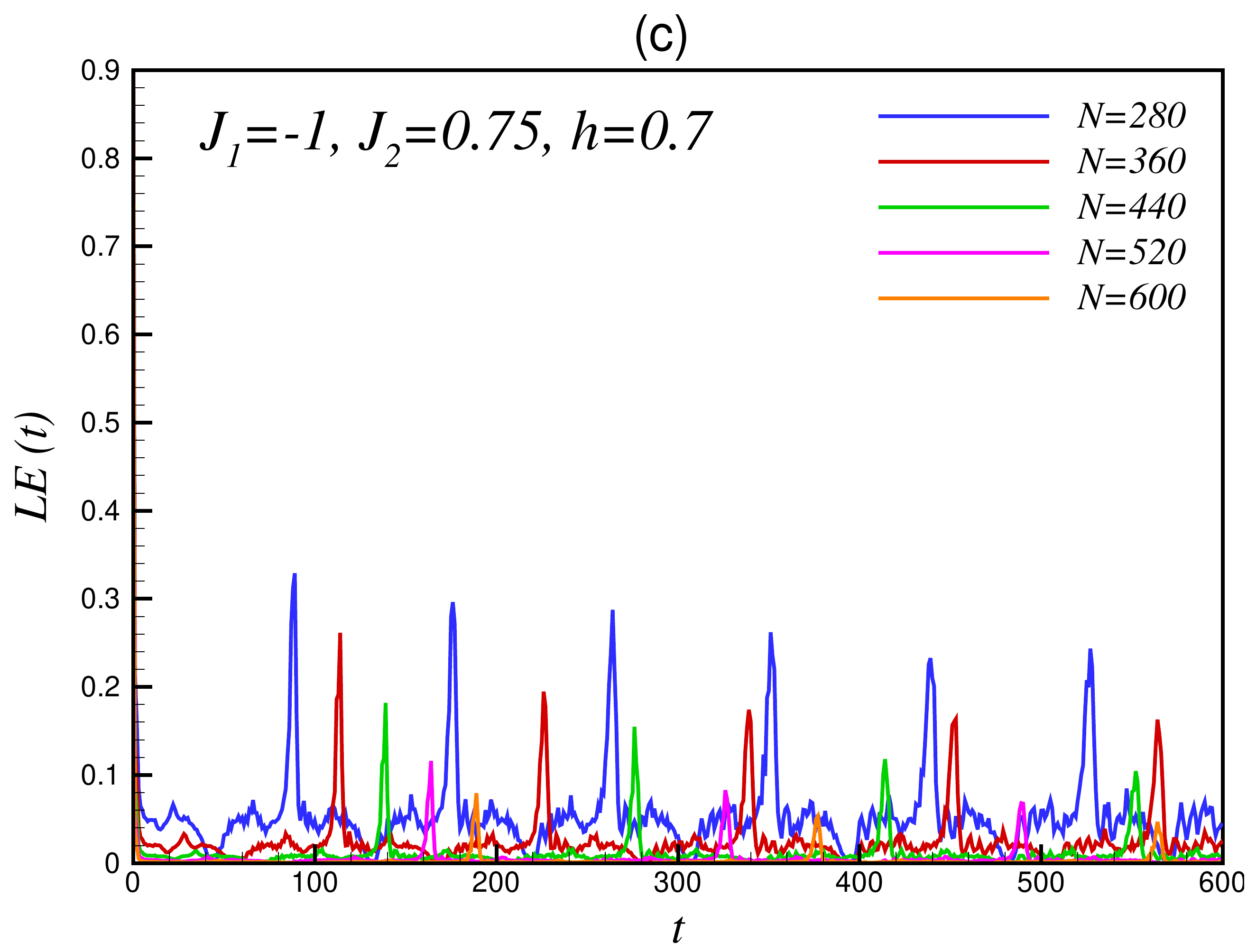}}
\caption{(Color online.) The LE starting from the different phases to
the critical points (a) $J_{1}=1, J_{2}=1.25, L_{1}=1, L_{2}=0, h=0.5$, (b) $J_{1}=1, J_{2}=1.25, h=1.5$,
(c) $J_{1}=-1, J_{2}=0.75, h=0.5$, for different system sizes.}
\label{fig10}
\end{figure*}

\begin{figure*}
\centerline{\includegraphics[width=6.2cm]{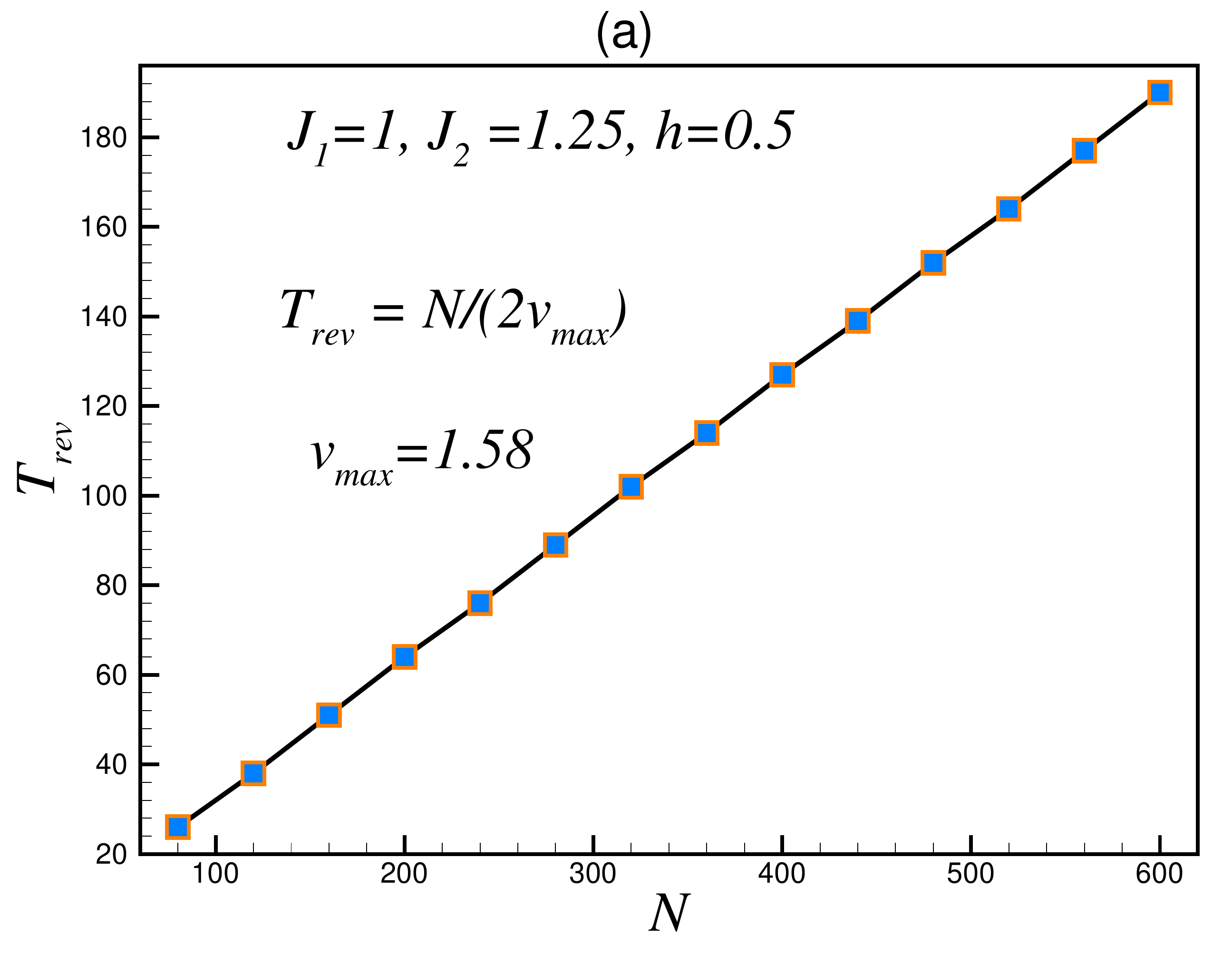}
\includegraphics[width=6.2cm]{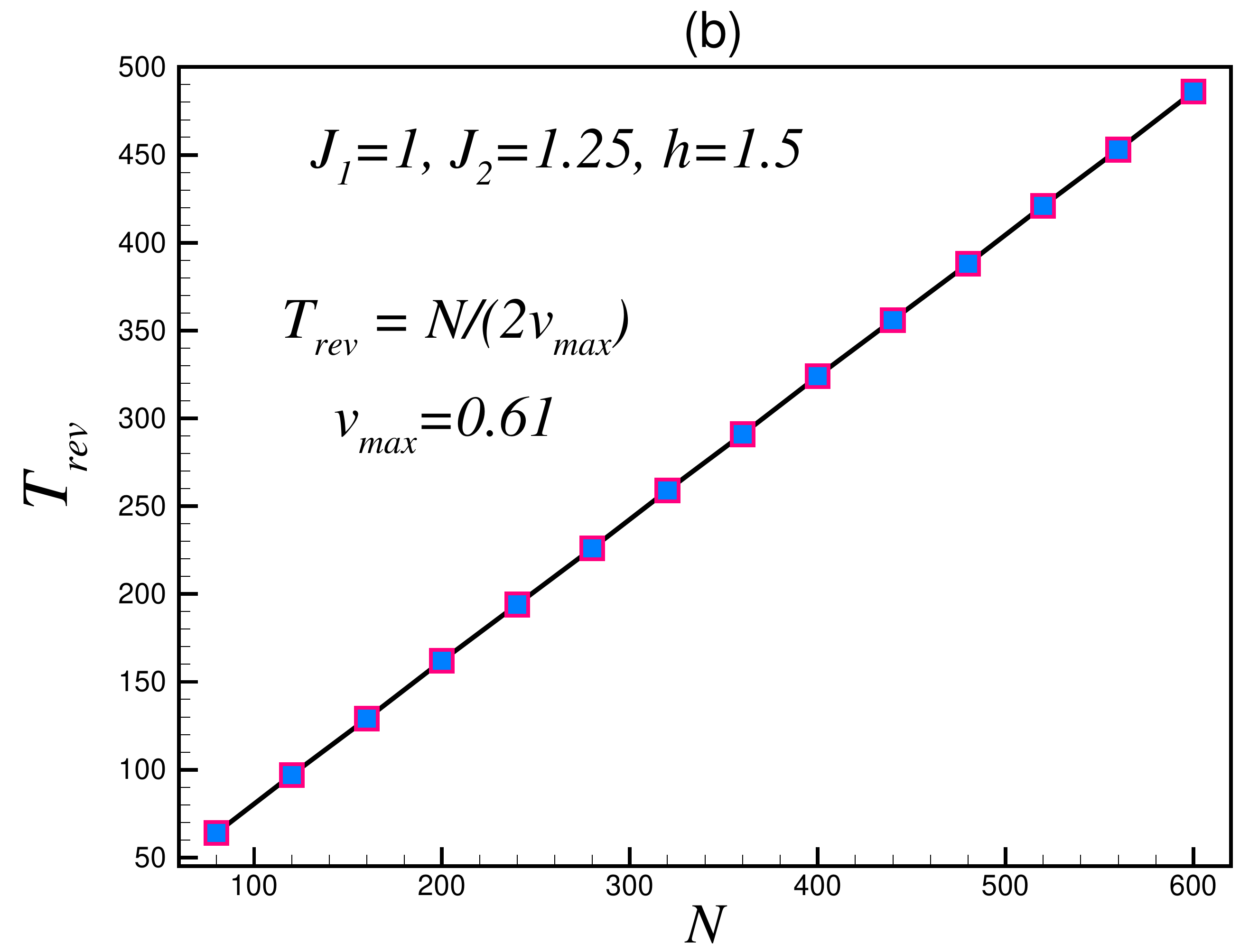}
\includegraphics[width=6.2cm]{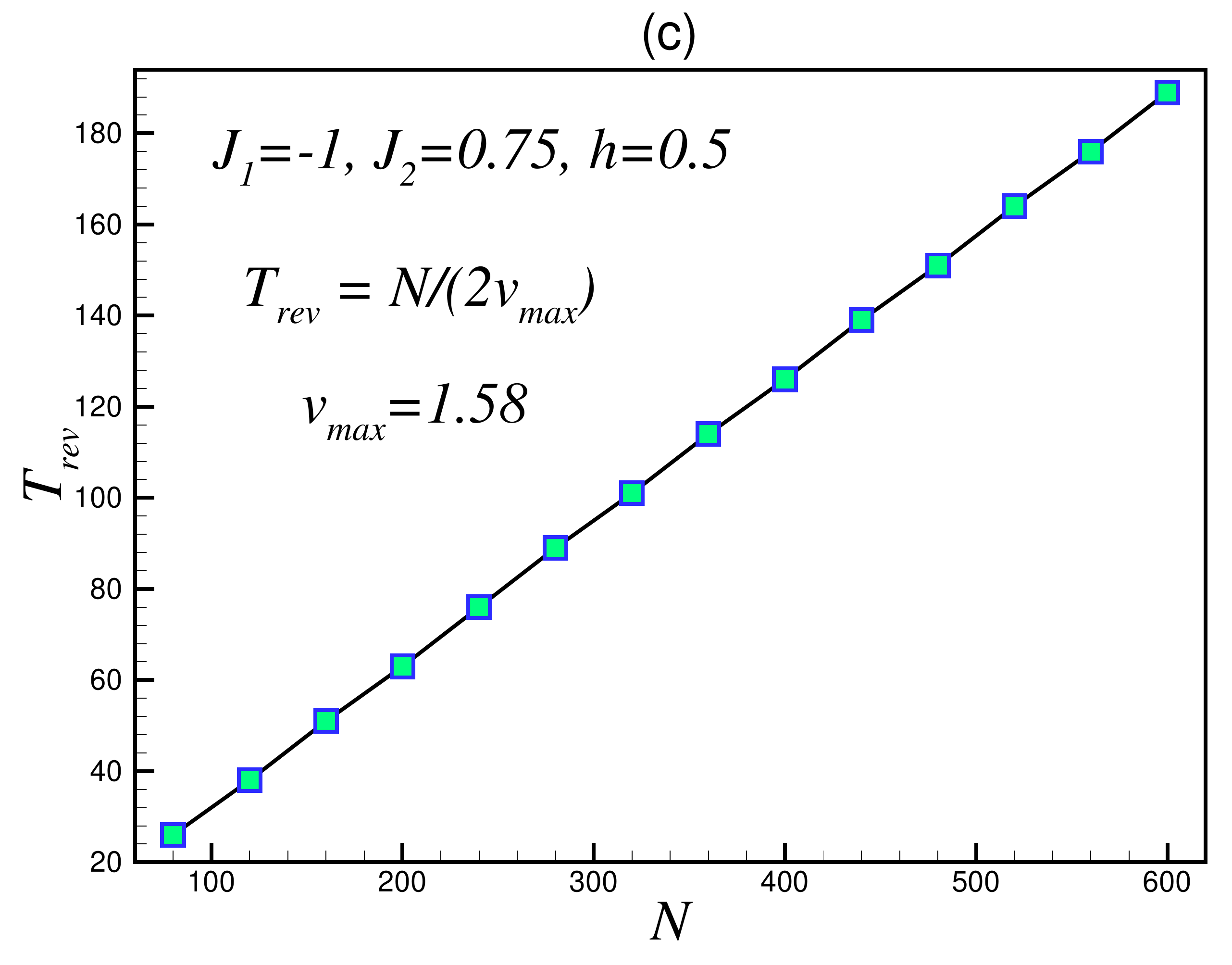}}
\caption{(Color online.) The scaling behavior of the first revival time $T_{rev}$
for different-length chains for a quenches at the different types of critical points
(a) $J_{1}=1, J_{2}=1.25, h=0.5$, (b) $J_{1}=1, J_{2}=1.25, L_{1}=1, L_{2}=0, h=1.5$, (c) $J_{1}=-1, J_{2}=0.75, h=0.5$ .}
\label{fig11}
\end{figure*}

\begin{figure*}
\centerline{\includegraphics[width=6.2cm]{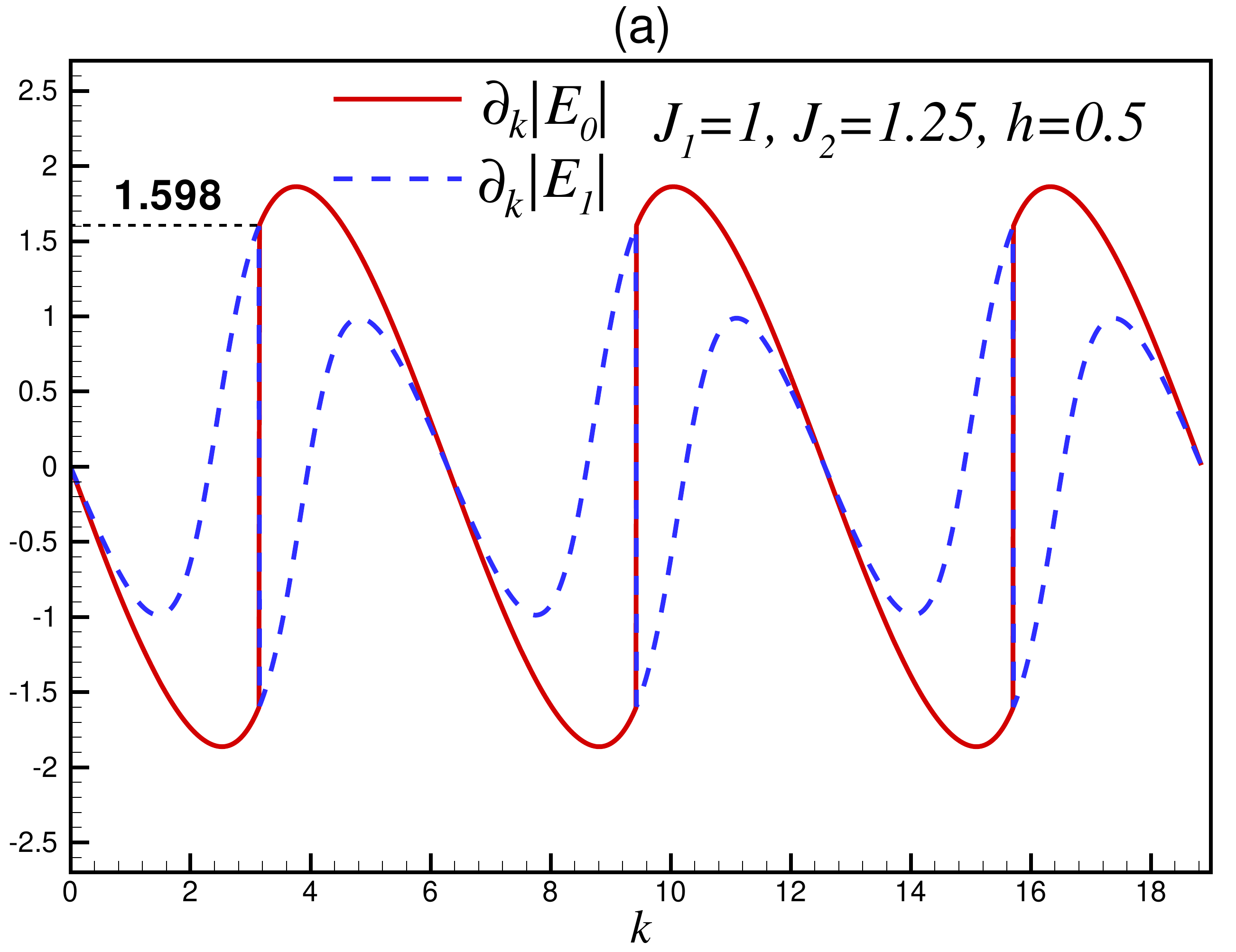}
\includegraphics[width=6.2cm]{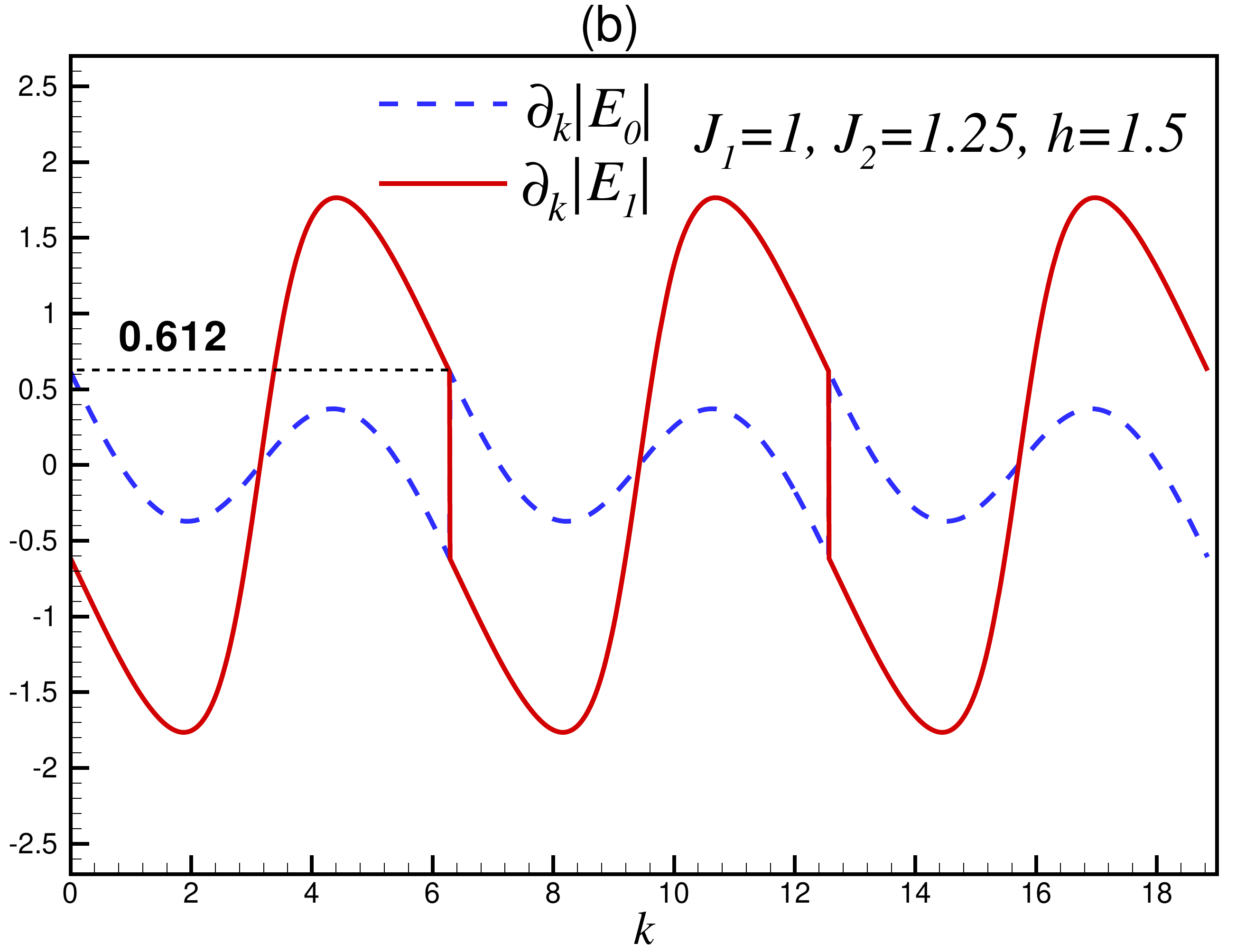}
\includegraphics[width=6.2cm]{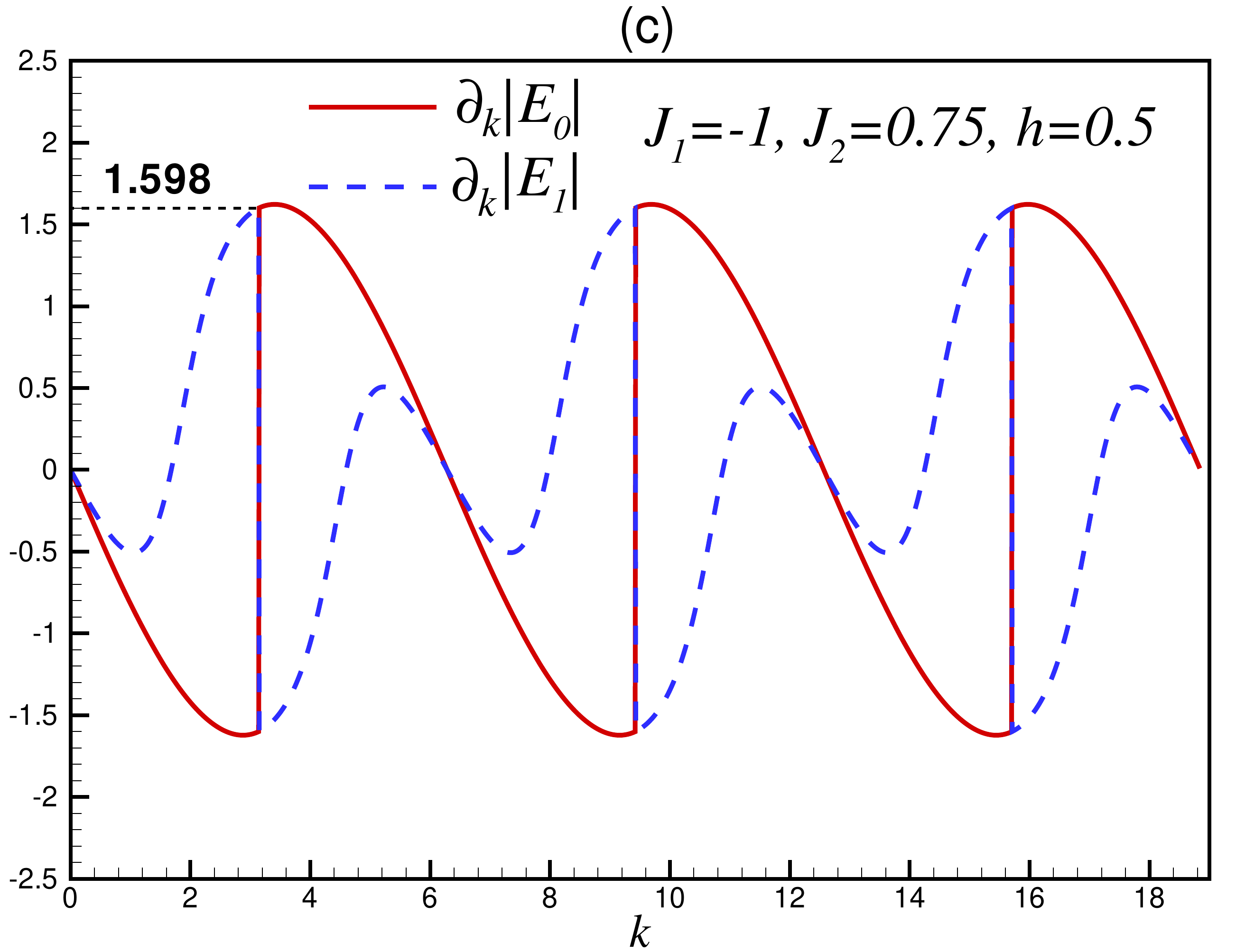}}
\caption{(Color online.) The first derivative of ground state ($E_{0}$) and excited state ($E_{1}$) with respect to the momentum $k$
 for quenches at the different critical points (a) $J_{1}=1, J_{2}=1.25, L_{1}=1, L_{2}=0, h=0.5$, (b) $J_{1}=1, J_{2}=1.25, h=1.5$, (c) $J_{1}=-1, J_{2}=0.75, h=0.5$
$h=0.1$, (b) $h=0.5$.}
\label{fig12}
\end{figure*}

We have plotted $\chi_{F}/N$ for $J_{1}=1$ and $J_{2}=2$
versus $h$ in Fig. \ref{fig5} (b) for different lattice sizes which shows the singular
behavior as the size of the system becomes large. As it manifests the divergences of
$\chi_{F}$ occurs at $h_{c_{1}}=1$ and $h_{c_{2}}=\sqrt{3}$.
The similar analysis shows the scaling behavior of the position of the first and second
maximum point ($h_{Max_{1}}, h_{Max_{2}}$) of $\chi_{F}$ tends toward the critical point like $h_{Max_{i}}=h_{c_{i}}+N^{-\theta_{i}}$ ($i=1,2$) and results are presented in Table. (I).
Moreover, our results show a linear behavior of $\chi_{F}|_{h_{Max_{i}}}$ versus $\ln(N)$
with the exponent $\tau_{i}$ (see Table I).
We illustrate the finite-size scaling behaviors of $\chi_{F}$ around its
maximum points. It shows that the fidelity susceptibility can be approximately
collapsed to a single curve. These results show that all the key ingredients
of the finite-size scaling are present in these cases too. In these cases
scaling is fulfilled with the critical exponent $\nu=1$ (see Table. (I)),
in agreement with the previous results and the universality hypothesis.
The behavior of the fidelity susceptibility has been also investigated in other regions.
Our calculations show that the non-analytic and scaling behavior of fidelity
susceptibility are the same as the former results with the same finite size scaling
(see Table I).
It would be worth to mention that the defect density $n_{ex}(h)$ can
be related to the fidelity susceptibility $\chi_{F}$ as $n_{ex}=(1/L^{d})h\chi_{F}(h)$,
where $L$ is the linear dimension of a $d$-dimensional quantum system.
The defect density can be defined for any Hamiltonian system, for a sudden quench
close to a quantum critical point \cite{Patel}. In a sudden quench, when a parameter $h$
in the Hamiltonian of the system is changed suddenly, the wave function of the
system does not have sufficient time to evolve. If the system
is initially prepared in the ground state for the initial value of
the driving parameter, it can not be in the ground state of the
final Hamiltonian. Consequently, there are defects in the final state and its scaling
follows from the scaling of fidelity susceptibility.

\section{Quench dynamics and Loschmidt echo}

Studying the quench dynamics of the systems can be done in several methods.
One of the possible scenarios is Kibble-Zurek mechanism \cite{Zurek} which
estimates the functional dependence of the density of defects on the quenching rate
in a system crossing a quantum critical point. Moreover, the Landau-Zener formula \cite{Suzuki}
turns out to be very valuable in investigating the transition probability to the excited
state of a two-level systems when a parameter in the Hamiltonian is changed at a
slow and uniform rate. Another formalism which has been applied in this section to
investigate the time evolution of the ground state after a critical quantum quench
is Loschmidt echo. It should be mention that, although the Landau-Zener formalism is a powerful
tool to study the quench dynamics of systems, it is not possible to deal with the Hamiltonian
of the model Eq. (\ref{eq2}) as a two-level system.

As has been mentioned in introduction, a quantum quench is a sudden change in
the Hamiltonian of the system. The quantum system is originally prepared
in the ground state $|\psi_{0}\rangle$ of $H(h^{(1)})$, and at time $t=0$
the parameters are switched to different values $h^{(2)}$. The system
then evolves unitarily with the quench Hamiltonian $H(h^{(2)})$
according to $|\psi_{0}(t)\rangle=U(t)|\psi_{0}\rangle$, where $U(t)=\exp(-iH(h^{(2)})t )$.
An important quantity describing the time evolution is the
Loschmidt echo (LE) defined as

\bea
\label{eq8}
LE(h^{(1)}, h^{(2)},t)=|\langle\psi_{0}(h^{(1)})|U(t)|\psi_{0}(h^{(1)})\rangle|^2,
\eea

which gives a measure of the distance between the time evolved
state $|\psi_{0}(t)\rangle$ and the initial state $|\psi_{0}\rangle$. High
values of the LE mean that the system is approaching the initial state
($LE(t=0)=1$).
The LE typically decays exponentially in a short time $T_{rel}$ (relaxation time) from 1 to
its average value, around which it oscillates.
Revivals are also visible in the LE as deviations from
the average value. The revivals have been defined as a time instances ($T_{rev}$)
at which the signal $LE(t)$ differs from the average value.
The structure of these revivals could be greatly affected
by criticality \cite{Quan}.
Since the model is exactly solvable and the whole spectrum and the eigenstate
of the model has been obtained, this leads to an exact
expression for the LE, and the revival times can be extracted
by inspecting its time dependence.
To obtain the analytical expression for LE, we should express the old ground state
$|\psi_{0} (h^{(1)})\rangle$ in terms of the eigenstates of the quench
Hamiltonian $H(h^{(2)})$. The time evolution in the EQC  model in a transverse field
after a quantum quench is given by:

\bea
\no
|\psi_{0}(t)\rangle=\frac{1}{\sqrt{N_{0}(h^{(1)})}}\Big[\sum_{m=0}^{7}
\frac{e^{-\imath E_{m}(h^{(2)})t}}{\sqrt{N_{m}(h^{(2)})}}
\sum_{j=1}^{8}v^{(0)}_{j}(h^{(1)})
(v^{(m)}_{j}(h^{(2)}))^{\ast}|\psi_{m}(h^{(2)})\rangle\Big],
\eea

where $N_{m}=\sqrt{|\langle\psi_{m}|\psi_{m}\rangle|},~ (m=0,\cdots, 8)$, and the $LE$ is thus

\bea
\no
LE(h^{(1)}, h^{(1)},t)=\Big|\frac{1}{N_{0}(h^{(1)})}\Big[\sum_{m=0}^{7}
\frac{e^{-\imath E_{m}(h^{(2)})t}}{N_{m}(h^{(2)})}
\Big(\sum_{j=1}^{8}v^{(0)}_{j}(h^{(1)})(v^{(m)}_{j}(h^{(2)}))^{\ast}\Big)
\Big(\sum_{n=1}^{8}(v^{(0)}_{j}(h^{(1)}))^{\ast}(v^{(m)}_{n}(h^{(2)}))\Big)\Big]
\Big|^{2}
\eea

In the following, the detailed analysis of the LE has been shown for
different types of quenches.
Three-dimensional panorama of the LE has been
plotted as a functions of a transverse field and $t$ in Fig. \ref{fig6}
for $J_{1}=1$, $J_{2}=1.25$ and system size set as $N=100$.
As one can see, LE experiences a sharp decay around the critical points
$h_{\pi}=0.5$ and $h_{0}=1.5$ which exactly correspond to the critical points that
have been obtained using the fidelity and gap analysis $h_{\pi}=\sqrt{(J_{1}+L_{2})(J_{2}-L_{1})},~h_{0}=\sqrt{(J_{1}+L_{2})(J_{2}+L_{1})}$.
As it is clear, the structure of revivals are different at the critical points and $T_{rev}(h_{0})/T_{rev}(h_{\pi})\simeq2.5$.

Consider the critical pint $J_{1}=1, J_{2}=1.25, h=0.5$. As we see in Fig. \ref{fig2} (b), this point lies between two different
phases. Increasing (decreasing) $J_{1}$ and  $J_{2}$ (magnetic field $h$) forces the system into the antiparallel ordering
of the $y$ spin component on even bonds phase and decreasing (increasing) $J_{1}$ and  $J_{2}$
(magnetic field $h$) drive the model to the spin-flop phase. In Fig. \ref{fig7} LE has been plotted for a different types of quenches.
For this critical point the behavior of the LE does not depend on which phase
the system is prepared in qualitatively, and $T_{rev}$ is the same for all the quenches.

In Fig. \ref{fig8} the system is quenched from different ground
states corresponding to different phases (saturate ferromagnetic phase and spin-flop
phase) to the critical point $J_{1}=1, J_{2}=1.25, h=1.5$.
The numerical simulations show that the details of the evolution
are somewhat different, but the structure of the revivals is the same
for all the quenches, and it oscillates around a relatively high mean
value in Fig. \ref{fig8} (b) in compare with those one in Figs. \ref{fig8} (a) and \ref{fig8} (c).
The behavior of the LE after a quench of $J_{1}$ and
magnetic field (Figs. \ref{fig8} (a) (c)) are similar to the
one obtained for critical quenches in the XY model \cite{Happola}.

We also quench to the critical point $J_{1}=-1, J_{2}=0.75, h=0.5$ which
lies between SAF phase and SF phase (Fig. (\ref{fig9})).
In this case, the details of the LE evolution are the same too, and the only difference
is their oscillations around a different mean values.
These indicate the universality of the revival structure in which the
initial state and the size of the quench are unimportant.
The question which is relevant in the measurement
process is how the revivals time ($T_{rev}$) can be derived in the LE?
We can address this question for the class of quasifree Fermi systems
to Refs. \cite{Happola, Montes} in which the phenomenon of the revivals after a
quantum quench is construed as a recombination of the fastest quasiparticles in the system.
Specifically, the speed of the fastest excitations in the system is upper bounded
by the Lieb-Robinson speed and this upper bound gives a lower bound to the revival time \cite{Montes}.
On the other hand, the most distinct revivals are given by the steady values of the
group velocity
$v_{g}(k)=|\frac{\partial (E_{k}^{q}+E_{k}^{p})}{\partial k}|=|\frac{\partial E_{0}}{\partial k}|$ \cite{Happola} and in this
scenario the first revival is the one corresponding to the maximum group velocity
($v_{max}=\max_{k} v_{g}(k)$) and the revival time scale could be given
by the following estimate

\bea
\label{eq9}
T_{rev}=\frac{N}{2v_{max}}.
\eea

To investigate the revival times in LE of the extended quantum compass model in a transverse field, we have
plotted the LE for various system sizes in Fig. \ref{fig10} for different types of quenches.
It is easy to see that increasing the length of the chain decreases the oscillation
of the LE and the height of first revival decreases gradually and, finally,
vanishes as $N\rightarrow\infty$.
Examining the details shows the scaling behavior of first revival time
$T_{rev}$ versus $N$. This is plotted in Fig. \ref{fig11}, which shows the linear behavior of $T_{rev}$
versus $N$. The scaling behavior is

\bea
\label{eq10}
T_{rev}=KN^{\delta},
\eea

with exponent $\delta=1$.

A similar analysis can be carried out to obtain the $K$ coefficient. Our calculations show that
$K=\frac{1}{2v_{max}}$ where $v_{max}$ are given by following estimate

\bea
\no
v_{max}=\min(|\frac{\partial (E_{k}^{q}\pm E_{k}^{p})}{\partial k}|_{max})
\label{eq11}
=\min(|\frac{\partial E_{0}}{\partial k}|_{max},|\frac{\partial E_{1}}{\partial k}|_{max})
\eea

In Fig. \ref{fig11} the first derivative of the ground state $E_{0}$ and excited state $E_{1}$ with respect to
the momentum $k$ has been depicted for different critical points.
The numerical calculations show that $v_{max}=\min(|\frac{\partial E_{0}}{\partial k}|_{max},|\frac{\partial E_{1}}{\partial k}|_{max})$
and the revival time shows the universality and scaling around the quantum critical points consistent with those predicted by Eq. (\ref{eq10}) and $T_{rev}(h_{0})/T_{rev}(h_{\pi})=v_{max}(h_{\pi})/v_{max}(h_{0})=2.61$ which consistent with
those obtained in Fig. \ref{fig6}. A surprising result occurs in the critical surfaces which are the boundaries between the spin-flop (I) and the saturated ferromagnetic (III) phases and between the strip antiferromagnetic (IV) and the saturated ferromagnetic (III) phases which correspond to $h_{\pi}=\sqrt{(J_{1}+L_{2})(J_{2}-L_{1})}$.
In this critical magnetic field, the revival time of the LE is inversely proportional to the the maximum group velocity of
the excited state quasiparticles, while in $h_{0}=\sqrt{(J_{1}+L_{2})(J_{2}+L_{1})}$ critical field
the revival time is proportional to the inverse of the maximum group velocity of the ground state quasiparticles.
It should be pointed out that the regions of phase space with a different Lieb-Robinson speed
do not correspond to different quantum phases. Quantum phase transition happens in the ground state at a
singular point of the phase space, whereas the Lieb-Robinson speed is the maximum speed of any
signal in any state and not just small excitations over the ground state \cite{Schwarz}.

\section{Summary and conclusions \label{conclusion}}
In this work we have studied the ground state fidelity, fidelity susceptibility,
and quench dynamics of the one-dimensional extended quantum compass model (EQCM) in a transverse field.
We use the Jordan-Wigner transformation to construct an explicit analytic expressions of the
ground state fidelity and Loschmidt echo (LE) of this model.
We show how the fidelity susceptibility and LE could
detect the quantum phase transitions in the inhomogeneous system.
Moreover, we have investigated the universality and scaling properties of
fidelity susceptibility and revival time. The results show that the fidelity susceptibility
exhibits beautiful scaling law close to the critical points with exponent $\nu=1$ exactly corresponds
to the correlation length exponent of the Ising model in a transverse field.

We show that the LE exhibits a universal structure of revivals that is independent of
the initial state and the size of the quench. The information travels through
the spin system via wave packets of quasiparticles, and the first
revival appears when the wave packets propagates with the
minimum of the group velocities of ground state and excited state.
This interpretation explain the universality of the revival structure since group velocities only depends
on the dispersion relation of the quasiparticles related to
the ground state and excited state of the quench Hamiltonian.
Depending on the critical surfaces, the structure of the revivals after critical quantum quenches
represents two different equilibration dynamics, whereas examination of the fidelity susceptibility shows the
same correlation length exponent for all critical points.

\begin{acknowledgments}
The author would like to thank G. Watanabe, A. Akbari and V. Karimipour for reading the manuscript and
valuable comments.
\end{acknowledgments}

\section{Appendix \label{Appendix}}

\subsection{Unitary Transformation}

The unitary transformation matrix $U$ which can transform the
Hamiltonian Eq. (\ref{eq1})into a diagonal form, has the following form

\bea \no U=
\left(%
\begin{array}{cccc}
  U_{1,E_{k}^{q}} & U_{2,E_{k}^{q}} & U_{3,E_{k}^{q}} & 1 \\
  U_{1,-E_{k}^{q}} & U_{2,-E_{k}^{q}} & U_{3,-E_{k}^{q}} & 1 \\
  U_{1,E_{k}^{p}} & U_{2,E_{k}^{p}} & U_{3,E_{k}^{p}} & 1 \\
  U_{1,-E_{k}^{p}} & U_{2,-E_{k}^{p}} & U_{3,-E_{k}^{p}} & 1 \\
\end{array}%
\right),
\eea

where

\begin{widetext}
\bea
\no
U_{1,\pm E_{k}^{\alpha}}&=&\frac{\frac{2h\pm E_{k}^{\alpha}}{J^{\ast}}-\frac{L^{\ast}}{J^{\ast}}\Big[\frac{L^{\ast}(2h\pm E_{k}^{\alpha})}{{L^\ast}^{2}-{J^\ast}^{2}}
-J^{\ast}\Big((2h\mp E_{k}^{\alpha})\Big(L({J^\ast}^{2}-{L^\ast}^{2})-L^{\ast}(2h\pm E_{k}^{\alpha})^{2}\Big)\Big)\Big]}
{\Big[{L^\ast}^{2}-{J^\ast}^{2}\Big]\Big[J^{\ast}(\gamma a-|L|^2)-J{L^\ast}^{2}\Big]},\\
\no
U_{2,\pm E_{k}^{\alpha}}&=&\frac{\Big[\frac{-L^{\ast}(2h\pm E_{k}^{\alpha})}{{L^\ast}^{2}-{J^\ast}^{2}}
+J^{\ast}\Big((2h\mp E_{k}^{\alpha})\Big(L({J^\ast}^{2}-{L^\ast}^{2})-L^{\ast}(2h\pm E_{k}^{\alpha})^{2}\Big)\Big)\Big]}
{\Big[{L^\ast}^{2}-{J^\ast}^{2}\Big]\Big[J^{\ast}(\gamma a-|L|^2)-J{L^\ast}^{2}\Big]},\\
\no
U_{3,\pm E_{k}^{\alpha}}&=&-\frac{\Big[L({J^\ast}^{2}-{L^\ast}^{2})-L^{\ast}(2h\pm E_{k}^{\alpha})^{2}\Big)\Big]}
{\Big[J^{\ast}(\gamma a-|L|^2)-J{L^\ast}^{2}\Big]},\\
\no
\eea
\end{widetext}

$\alpha=q, p$, $\gamma=1$ for $\alpha=q$ and $\gamma=-1$ for $\alpha=p$.

\subsection{Ground State}

By using the unitary transformation the unnormalized eigenvectors and eigenvalues have the following expressions in
terms of the coupling constants:

\begin{widetext}
\bea
\no
|\psi_{m}\rangle=\prod_{k}&[&v_{1}^{m}|0\rangle+v_{2}^{m}~c_{k}^{{q\dag}}c_{-k}^{{q\dag}}|0\rangle+
v_{3}^{m}~c_{k}^{{q\dag}}c_{-k}^{{p\dag}}|0\rangle+v_{4}^{m}~c_{-k}^{{q\dag}}c_{k}^{{p\dag}}|0\rangle
+v_{5}^{m}~c_{k}^{{p\dag}}c_{-k}^{{p\dag}}|0\rangle+v_{6}^{m}~c_{k}^{{q\dag}}c_{k}^{{p\dag}}|0\rangle\\
\no
&+&v_{7}^{m}~c_{-k}^{{q\dag}}c_{-k}^{{p\dag}}|0\rangle+v_{8}^{m}~c_{k}^{{q\dag}}c_{-k}^{{q\dag}}c_{k}^{{p\dag}}c_{-k}^{{p\dag}}|0\rangle],~m=0,1,\cdots,7.
\eea
\end{widetext}

\begin{widetext}
\bea
\no
E_{0}&=&-\sqrt{2}\sqrt{4h^{2}+|J|^{2}+|L|^{2}+C},\\
\no
v_{1}^{0}&=&\frac{64h^{4}-{J^\ast}^{2}L^{2}+J^{2}(2{J^\ast}^{2}-{L^\ast}^{2})+2|J|^{2}C-16E_{0}h^{3}
-4h(|J|^{2}-|L|^{2}+C)E_{0}+8h^{2}(3|J|^{2}-2|L|^{2}+2C)}
{8h^{2}|J|^{2}+2|J|^{2}C+2|J|^{4}-J^{2}{L^\ast}^{2}-{J^\ast}^{2}L^{2}},\\
\no
v_{2}^{0}&=&v_{5}^{0}=-\frac{E_{0}(4h-E_{0})(LJ^{\ast}-JL^{\ast})/2}
{8h^{2}|J|^{2}+2|J|^{2}C+2|J|^{4}-J^{2}{L^\ast}^{2}-{J^\ast}^{2}L^{2}},\\
\no
v_{3}^{0}&=&-\frac{(4h-E_{0})(4h^{2}J+{J^\ast}(J^{2}-L^{2})+JC)}
{8h^{2}|J|^{2}+2|J|^{2}C+2|J|^{4}-J^{2}{L^\ast}^{2}-{J^\ast}^{2}L^{2}},\\
\no
v_{4}^{0}&=&-\frac{(4h-E_{0})(4h^{2}{J^\ast}+J({J^\ast}^{2}-{L^\ast}^{2})+{J^\ast}C)}
{8h^{2}|J|^{2}+2|J|^{2}C+2|J|^{4}-J^{2}{L^\ast}^{2}-{J^\ast}^{2}L^{2}},\\
\no
v_{6}^{0}&=&~v_{7}^{0}=0,~v_{8}^{0}=1,
\eea
\end{widetext}


\begin{widetext}
\bea
\no
E_{1}&=&-\sqrt{2}\sqrt{4h^{2}+|J|^{2}+|L|^{2}-C},\\
\no
v_{1}^{1}&=&\frac{-64h^{4}+{J^\ast}^{2}L^{2}-J^{2}(2{J^\ast}^{2}-{L^\ast}^{2})+2|J|^{2}C+16E_{1}h^{3}
+4h(|J|^{2}-|L|^{2}-C)E_{1}-8h^{2}(3|J|^{2}-2|L|^{2}-2C)}
{-8h^{2}|J|^{2}+2|J|^{2}C-2|J|^{4}+J^{2}{L^\ast}^{2}+{J^\ast}^{2}L^{2}},\\
\no
v_{2}^{1}&=&v_{5}^{1}=-\frac{E_{1}(4h-E_{1})(LJ^{\ast}-JL^{\ast})/2}
{-8h^{2}|J|^{2}+2|J|^{2}C-2|J|^{4}+J^{2}{L^\ast}^{2}+{J^\ast}^{2}L^{2}},\\
\no
v_{3}^{1}&=&-\frac{(4h-E_{0})(4h^{2}J+{J^\ast}(J^{2}-L^{2})-JC)}
{-8h^{2}|J|^{2}+2|J|^{2}C-2|J|^{4}+J^{2}{L^\ast}^{2}+{J^\ast}^{2}L^{2}},\\
\no
v_{4}^{1}&=&-\frac{(4h-E_{0})(4h^{2}{J^\ast}+J({J^\ast}^{2}-{L^\ast}^{2})-{J^\ast}C)}
{-8h^{2}|J|^{2}+2|J|^{2}C-2|J|^{4}+J^{2}{L^\ast}^{2}+{J^\ast}^{2}L^{2}},\\
\no
v_{6}^{1}&=&~v_{7}^{1}=0,~v_{8}^{1}=1,
\eea
\end{widetext}


\bea
\no
E_{2}=0,~v_{2}^{1}=v_{2}^{3}=v_{2}^{4}=v_{2}^{6}=v_{2}^{7}=v_{2}^{8}=0,v_{2}^{2}=-v_{2}^{5}=-1,
\eea


\bea
\no
E_{3}=0,~v_{3}^{1}=v_{3}^{2}=v_{3}^{3}=v_{3}^{4}=v_{3}^{5}=v_{3}^{7}=v_{3}^{8}=0,v_{3}^{6}=1,
\eea


\bea
\no
E_{4}=0,~v_{4}^{1}=v_{4}^{2}=v_{4}^{3}=v_{4}^{4}=v_{4}^{5}=v_{4}^{6}=v_{4}^{8}=0,v_{4}^{7}=1,
\eea


\bea
\no
E_{5}=0,~v_{5}^{2}&=&v_{5}^{5}=v_{5}^{6}=v_{5}^{7}=0,~v_{5}^{1}=-v_{5}^{8}=-1,\\
\no
v_{5}^{3}&=&\frac{4hL}{L{J^\ast}+J{L^\ast}},~v_{5}^{4}=\frac{4h{L^\ast}}{L{J^\ast}+J{L^\ast}},
\eea


\begin{widetext}
\bea
\no
E_{6}&=&\sqrt{2}\sqrt{4h^{2}+|J|^{2}+|L|^{2}-C},\\
\no
v_{1}^{6}&=&\frac{-64h^{4}+{J^\ast}^{2}L^{2}-J^{2}(2{J^\ast}^{2}-{L^\ast}^{2})+2|J|^{2}C+16E_{6}h^{3}
+4h(|J|^{2}-|L|^{2}-C)E_{6}-8h^{2}(3|J|^{2}-2|L|^{2}-2C)}
{-8h^{2}|J|^{2}+2|J|^{2}C-2|J|^{4}+J^{2}{L^\ast}^{2}+{J^\ast}^{2}L^{2}},\\
\no
v_{2}^{6}&=&v_{5}^{6}=-\frac{E_{6}(4h-E_{6})(LJ^{\ast}-JL^{\ast})/2}
{-8h^{2}|J|^{2}+2|J|^{2}C-2|J|^{4}+J^{2}{L^\ast}^{2}+{J^\ast}^{2}L^{2}},\\
\no
v_{3}^{6}&=&\frac{(4h-E_{6})(4h^{2}J+{J^\ast}(J^{2}-L^{2})-JC)}
{-8h^{2}|J|^{2}+2|J|^{2}C-2|J|^{4}+J^{2}{L^\ast}^{2}+{J^\ast}^{2}L^{2}},\\
\no
v_{4}^{6}&=&\frac{(4h-E_{6})(4h^{2}{J^\ast}+J({J^\ast}^{2}-{L^\ast}^{2})-{J^\ast}C)}
{-8h^{2}|J|^{2}+2|J|^{2}C-2|J|^{4}+J^{2}{L^\ast}^{2}+{J^\ast}^{2}L^{2}},\\
\no
v_{6}^{6}&=&~v_{7}^{6}=0,~v_{8}^{6}=1,
\eea
\end{widetext}


\begin{widetext}
\bea
\no
E_{7}&=&\sqrt{2}\sqrt{4h^{2}+|J|^{2}+|L|^{2}+C}\\
\no
v_{1}^{7}&=&\frac{64h^{4}-{J^\ast}^{2}L^{2}+J^{2}(2{J^\ast}^{2}-{L^\ast}^{2})+2|J|^{2}C-16E_{7}h^{3}
-4h(|J|^{2}-|L|^{2}+C)E_{7}+8h^{2}(3|J|^{2}-2|L|^{2}+2C)}
{8h^{2}|J|^{2}+2|J|^{2}C+2|J|^{4}-J^{2}{L^\ast}^{2}-{J^\ast}^{2}L^{2}},\\
\no
v_{2}^{7}&=&v_{5}^{7}=-\frac{E_{7}(4h-E_{7})(LJ^{\ast}-JL^{\ast})/2}
{-8h^{2}|J|^{2}-2|J|^{2}C-2|J|^{4}+J^{2}{L^\ast}^{2}+{J^\ast}^{2}L^{2}},\\
\no
v_{3}^{7}&=&-\frac{(4h-E_{7})(4h^{2}J+{J^\ast}(J^{2}-L^{2})+JC)}
{8h^{2}|J|^{2}+2|J|^{2}C+2|J|^{4}-J^{2}{L^\ast}^{2}-{J^\ast}^{2}L^{2}},\\
\no
v_{4}^{7}&=&-\frac{(4h-E_{7})(4h^{2}{J^\ast}+J({J^\ast}^{2}-{L^\ast}^{2})+{J^\ast}C)}
{8h^{2}|J|^{2}+2|J|^{2}C+2|J|^{4}-J^{2}{L^\ast}^{2}-{J^\ast}^{2}L^{2}},\\
\no
v_{6}^{7}&=&~v_{7}^{7}=0,~v_{8}^{7}=1,
\eea
\end{widetext}

in which $C=\sqrt{a^{2}-b}$.


\section*{References}
\begin{thebibliography}{99}

\bibitem{Osterloh}
A. Osterloh, Luigi Amico, G. Falci, and Rosario Fazio, Nature
(London) \textbf{416}, 608 (2002).

\bibitem{Vidal}
G. Vidal, J. I. Latorre, E. Rico, and A. Kitaev, Phys. Rev. Lett.
\textbf{90}, 227902 (2003).


\bibitem{Osborne}
T. J. Osborne, M. A. Nielsen, Phys. Rev. A. \textbf{66}, 032110
(2002); G. Vidal, J. I. Latorre, E. Rico, and A. Kitaev,
Phys. Rev. Lett. \textbf{90}, 227902 (2003); Y. Chen,
P. Zanardi, Z. D. Wang, F. C. Zhang, New J. Phys. 8,
\textbf{97} (2006); L. A. Wu, M. S. Sarandy, D.
A. Lidar, Phys. Rev. Lett. \textbf{93}, 250404 (2004).

\bibitem{kargarian1}
M. Kargarian, R. Jafari and A. Langari, Phys. Rev. A \textbf{76},
060304(R) (2007).

\bibitem{kargarian2}
M. Kargarian, R. Jafari, and A. Langari, Phys. Rev. A \textbf{77},
032346 (2008).

\bibitem{Jafari1}
R. Jafari, M. Kargarian, A. Langari, and M. Siahatgar, Phys. Rev.
B \textbf{78}, 214414 (2008).

\bibitem{kargarian3}
M. Kargarian, R. Jafari and A. Langari, Phys. Rev. A \textbf{79},
042319 (2009).

\bibitem{Jafari2}
R. Jafari, Phys. Rev. A \textbf{82}, 052317 (2010).

\bibitem{Langari}
A. Langari, and A. T. Rezakhani, New J. Phys. \textbf{14} 053014 (2012);
N. Amiri, and A. Langari, Phys. Status Solidi B. \textbf{250}, 537 (2013).

\bibitem{Jafari}
R. Jafari and A. Langari, Int. J. Quantum Inf. \textbf{9} (04), 1057, (2011).

\bibitem{Mehran}
E. Mehran, S. Mahdavifar, R. Jafari, Phys. Rev. A. \textbf{89}, 049903, (2014).


\bibitem{Jafari5}
R. Jafari, S. Mahdavifar, Prog. Theor. Exp. Phys. \textbf{4}, 043I02 (2014).


\bibitem{Zanardi}
P. Zanardi, P Giorda, and M. Cozzini, Phys. Rev. Lett.
\textbf{99}, 100603 (2007); P. Zanardi and N. Paunkovic,
Phys. Rev. E \textbf{74}, 031123 (2006).


\bibitem{Greiner}
M. Greiner, O. Mandel, T. W. H\"{a}nsch, and I. Bloch, Nature
(London) \textbf{415}, 39 (2002).

\bibitem{Jaksch}
D. Jaksch, C. Bruder, J. I. Cirac, C. W. Gardiner, and P. Zoller,
Phys. Rev. Lett. \textbf{81}, 3108 (1998).

\bibitem{Stoferle}
T. St\"{o}ferle, H. Moritz, C. Schori, M. Köhl, and T. Esslinger,
Phys. Rev. Lett. \textbf{92}, 130403 (2004).

\bibitem{Chandra}
Quantum Quenching, Annealing and Computation, edited by
A. K. Chandra, A. Das, and B. K. Chakrabarti (Springer,
Heidelberg, 2010).


\bibitem{Gorin}
T. Gorin, T. Prosen, T. H. Seligman, and M. Znidaric, Phys. Rep.
\textbf{435}, 33 (2006).

\bibitem{Jacquod}
Ph. Jacquod and C. Petitjean, Adv. Phys. \textbf{58}, 67 (2009).


\bibitem{Venuti2}
L. Campos Venuti, N. T. Jacobson, S. Santra, and P. Zanardi,
Phys. Rev. Lett. \textbf{107}, 010403 (2011).

\bibitem{Quan}
H. T. Quan, Z. Song, X. F. Liu, P. Zanardi, and C. P. Sun, Phys.
Rev. Lett. \textbf{96}, 140604 (2006).

\bibitem{Montes}
S. Montes, and A. Hamma, Phys. Rev. E. \textbf{86}, 021101 (2012).

\bibitem{Eriksson}
E. Eriksson and H. Johannesson, Phys. Rev. B \textbf{79}, 224424
(2009).

\bibitem{Mahdavifar}
S. Mahdavifar, Eur. Phys. J. B \textbf{77}, 77 (2010).



\bibitem{Motamedifar}
M. Motamedifar, S. Mahdavifar, and S. F. Shayesteh, Eur.
Phys. J. B \textbf{83}, 181 (2011).


\bibitem{Jafari3}
R. Jafari, Phys. Rev. B \textbf{84}, 035112 (2011).


\bibitem{Jafari4}
R. Jafari, Eur. Phys. J. B \textbf{85}, 167  (2012).


\bibitem{You}
W. L. You, Eur. Phys. J. B \textbf{85}, 83 (2012).


\bibitem{Liu}
G.H. Liu, W. Li, and W. L. You, Eur. Phys. J. B \textbf{85}, 168 (2012).


\bibitem{Motamedifar2}
M. Motamedifar, S. Nemati, S. Mahdavifar, and S. F. Shayesteh,
Phys. Scr. \textbf{88}, 015003 (2013).


\bibitem{Happola}
J. H\"{a}pp\"{o}l\"{a}, G. B. Hal\'{a}sz, and A. Hamma, Phys. Rev. A \textbf{85},
032114 (2012).


\bibitem{Vidal2}
G. Vidal, Phys. Rev. Lett. \textbf{98}, 070201 (2007).



\bibitem{Jordan}
E. Lieb, T. Schultz, and D. Mattis, Ann. Phys. (N.Y.) \textbf{16}, 407 (1961);
E. Barouch and B. M. McCoy, Phys. Rev. A \textbf{3}, 786 (1971);
J. B. Kogut, Rev. Mod. Phys. \textbf{51}, 659 (1979);
J. E. Bunder and R. H. McKenzie, Phys. Rev. B \textbf{60}, 344 (1999).


\bibitem{Perk}
J. H. H. Perk, H. W. Capel, and M. J. Zuilhof, Physica \textbf{81A} 319 (1975).



\bibitem{Sengupta}
K. Sengupta, D. Sen, and S. Mondal, Phys. Rev. Lett. \textbf{100}, 077204 (2008);
S. Mondal, D. Sen, and K. Sengupta, Phys. Rev. B \textbf{78}, 045101 (2008).

\bibitem{You2}
Wen-Long You, Ying-Wai Li, and Shi-Jian Gu, Phys. Rev. E \textbf{76}, 022101 (2007).

\bibitem{Venuti}
L. Campos Venuti and P. Zanardi, Phys. Rev. Lett. \textbf{99},
095701 (2007).

\bibitem{Hamma}
A. Hamma, e-print: quant-ph/0602091.

\bibitem{Jafari6}
R. Jafari, Phys. Lett. A \textbf{377},3279 (2013).


\bibitem{Patel}
Aavishkar A. Patel and Amit Dutta, Phys. Rev. B. \textbf{86}, 174306 (2012).


\bibitem{Zurek}
W. H. Zurek, U. Dorner, and P. Zoller, Phys. Rev. Lett. \textbf{95}, 105701 (2005).

\bibitem{Suzuki}
S. Suzuki and M. Okada, \textit{in Quantum annealing and related optimization methods}
edited by A Das and B K Chakrabarti (Springer-Verlag, Berlin, 2005) p. 185.




\bibitem{Schwarz}
I. Pr\'{e}mont-Schwarz, and J. Hnybida, Phys. Rev. A. \textbf{81},
062107 (2010).

\end{thebibliography}
\end{document}